\documentclass[dvipdfmx]{PoS}
\usepackage{wrapfig}
\usepackage{lineno}
\usepackage{placeins}
\usepackage{graphicx}
\usepackage{color}
\title{Neutrino Astronomy (Rapporteur Talk)}

\ShortTitle{Neutrino Astronomy}

\author{Aya Ishihara$^{1}$ 
\\
{\it 
$^1$Department of Physics, Graduate School of Science, Chiba University, Chiba 263-8522, Japan \\
}

E-mail: \email{aya@hepburn.s.chiba-u.ac.jp}
}

\abstract{This report is the write-up of a rapporteur talk on neutrino astronomy given at the 34th International Cosmic Ray Conference in The Hague, Netherlands, in 2015.
	Here, selected contributions on the neutrino astronomy from the total of 40 talks and 90 posters presented in NU sessions at the 34th ICRC are summarized in the attempt of providing a status report on this rapidly glowing new field.
	The field of neutrino astronomy has recently experienced a "phase 
transition" since the first observation of high energy cosmic neutrinos.
Extensive efforts have been made to identify the origin of the 
neutrino flux observed in the 100~TeV to PeV region, from both theoretical and experimental perspectives. 
In addition, the search for neutrino fluxes beyond the observed level 
has become increasingly important 
for further understanding the origin of the observed cosmic-ray up to $10^{20}$~eV.
Although the IceCube Neutrino Observatory is the only experiment currently measuring this neutrino flux, 
its initial measurements have been confirmed via analysis using several
independent detection channels.
Further, there have been a number of developments in the search for neutrino point sources, 
while no successful observations have yet been reported. 
Following the IceCube observations, a large number of 
studies of next-generation neutrino detectors, including 
up-scaled 
underground Cherenkov neutrino detectors and Cherenkov radio 
neutrino detectors, have been reported. 
}

\FullConference{The 34th International Cosmic Ray Conference,\\
		30 July- 6 August, 2015\\
		The Hague, The Netherlands}

\begin{document}
\section{Introduction}

In 1987, the detection of a transient high-level flux of supernova neutrinos with energies of
a few tens of MeV marked the beginning of neutrino astronomy \cite{sn1987a}.
A quarter of a century later, in 2012, the IceCube Neutrino Observatory reported the detection of two very high-energy neutrinos, which had energies of approximately one PeV \cite{pev}. 
This value is roughly one hundred million times greater than the energies of the previously observed 
cosmic neutrinos from a supernova. The discovery of two PeV neutrinos implied the existence of an astrophysical 
neutrino flux beyond the atmospheric neutrino background in the high-energy region.
Follow-up observations further indicated a steady flux of extraterrestrial neutrinos in the energy region from a few tens of TeV up to PeV \cite{hese2year}.
Increased data acquisition and statistical analyses using neary independent channels and samples confirmed their significance \cite{mese, numu, hese3year}; 
thus, it was declared that neutrino astronomy was moving into a new phase.
These extraterrestrial neutrinos, which are undeflected
in the galactic and extra-galactic magnetic fields and unattenuated in the photon-filled universe,
are expected to be a key tool as regards identifying ultra-high-energy (UHE) cosmic accelerators.
This is a considerable success for the astroparticle physics community and for all those who committed to designing and constructing the largest neutrino 
detector on Earth (including IceCube, ANTARES, NT-200, AMANDA, DUMAND etc.).

The IceCube collaboration 
presented evidence of this astrophysical neutrino flux just two years ago, at International Cosmic Ray Conference (ICRC) 2013 in Rio de Janeiro \cite{icrc2013}.
The measured neutrino flux level is close to the theoretical upper bound inferred from the ultra-high-energy cosmic ray (UHECR) flux 
measurements \cite{waxman}. 
%
%
%
%
At present, we must question the origin of this observed neutrino flux.
For experiments with a limited number of detectable events, 
this rather complex question can be interpreted as a 
composition of four simpler questions, i.e., whether or not the flux is isotropic; 
whether or not the flux is equal in neutrino flavor ratio; whether the flux shape follows a simple power-law; or, 
if not, how the flux shape can be modeled.
The answers to these simple questions are expected to be accessible based on the results of ongoing experiments, and
will constitute the basis for constraining and constructing future source models.
The spectral index and break/cut-off energy indicate cosmic-ray spectra in source objects at 
the mean energy region of approximately twenty times that of the observed neutrino energies.
The observed neutrino flux should give hints on, for instance, cosmic-ray interactions with the ambient photon field or matter in the source objects, 
the maximal acceleration energy that the object can be responsible for, or transitions of classes of sources account for the neutrino flux.

Moreover, there is another important question we wish to address, i.e., whether any other classes of 
neutrino flux exist besides those currently being observed.
As we know that there is a cosmic-ray flux that extends above $10^{20}$~eV \cite{nagano}, there is good reason to believe 
that a neutrino flux counterpart to the UHECR flux also exists, in the energy region higher than a few PeV \cite{berezinsky69}. 
To answer this question, it is likely that new Cherenkov neutrino detectors up-scaled by a 
factor of 5--10 are required.
For even higher energies which could exceed EeV energy region, it may be more practical to detect neutrino-induced signals 
using detectors that can be constructed with sparser instrumentation in a cost-effective manner. 
The use of radio Cherenkov detectors to search for coherent radio
Cherenkov signals produced by the Askaryan effect in a large volume 
of ice is a key concept for analysis of the neutrino energy region above 10~PeV. 

At ICRC 2015, 120 presentations were given in neutrino astronomy (NU) oral and poster sessions, with additional related contributions presented in cosmic-ray (CR) and gamma-ray (GR) astronomy sessions in the context of multi-messenger astronomy. 
In this write-up, only a limited number of experimental highlights from the NU sessions are summarized.

\section{Diffuse flux measurements}
Diffuse neutrino fluxes are considered to be superpositions of fluxes from individual 
neutrino point sources positioned close to the Earth or even at a very distant part of the universe.
While the level of neutrino flux of each point source is highly uncertain, 
the superposition of fluxes from all the astrophysical sources could create a 
detectable signal above those of atmospheric muons and neutrinos, which are due to the decays of protons, kaons, or short-lived charmed mesons in the atmosphere. 

Currently, the only detector capable of detecting steady beams of diffuse cosmic neutrinos 
is the IceCube Neutrino Observatory, which is the largest Cherenkov neutrino detector yet built \cite{gaisser2014}.
IceCube is a cubic-kilometer-scale deep-underground Cherenkov neutrino detector. 
The IceCube array comprises 5160 optical sensors
on 86 cables, over a 1-km$^3$ fiducial volume of ice at a depth of 1450--2450~m.
Construction of the IceCube detector was completed in December 2010. 
In 2008--2009, 2009--2010, and 2010--2011, 40,
59, and 79 of the total 86 cables, respectively, were deployed and successfully recording data with approximate fiducial volumes of
0.5, 0.7 and 0.9 km$^3$, respectively.
IceCube, which is located at the geographic South Pole, has been in 
full operation since May 2011.

The IceCube Neutrino Observatory maps the arrival times and positions of Cherenkov photo-electron signals 
using a simple three-dimensional (3D) array of optical sensor modules composed of photo-multiplier 
tubes and electrical components stored in a pressure-resistant spherical glass housing.
The recorded event-wise distributions of the Cherenkov light signals yield distinguishable features of the
parent charged particles.
The track-like features of the Cherenkov light signals are the signatures of muons propagating through the detector.
These muons are created by $\nu_\mu$ charged-current (CC) interactions outside or inside the IceCube 
instrumented volume. Similarly, track-like events from tau particles induced by  $\nu_\tau$ CC interactions
can be observed in the $\geq 10$-PeV energy range.
The particle showers produced at the neutrino interaction
vertex position are observed as ``cascade'' events.
A typical cascade-like feature of a photon distribution is a light flash 
propagating in all directions with a defined center.
Cascade events are induced by neutral current
(NC) interactions between all three neutrino flavors, as well as electron neutrino CC interactions. Although the cascade
channel is limited to the interactions occurring inside or near the IceCube detector, the corresponding energy
deposits in the detector are typically larger than those of muons or taus with equal energy.
With more precise knowledge of the properties of ice, which can be obtained using artificially calibrated light sources, 
IceCube will be capable of capturing more subtle features. The resultant complementary samples, which would be sensitive 
to different neutrino flavors and interactions, would enable us to study further details of neutrino spectra, such as the flavor compositions.

At ICRC 2015, the results of diffuse neutrino measurements from four different channels at IceCube including 
the results of a search for tau neutrinos \cite{tausearch} were presented.  
Although no tau neutrino event candidate was observed in the three years' worth of IceCube data considered in the reported study, the findings indicate good future prospects.

\subsection{Upward-going muon analysis}
\begin{figure*}[!h]
\centering
  \includegraphics[height=2.8in]{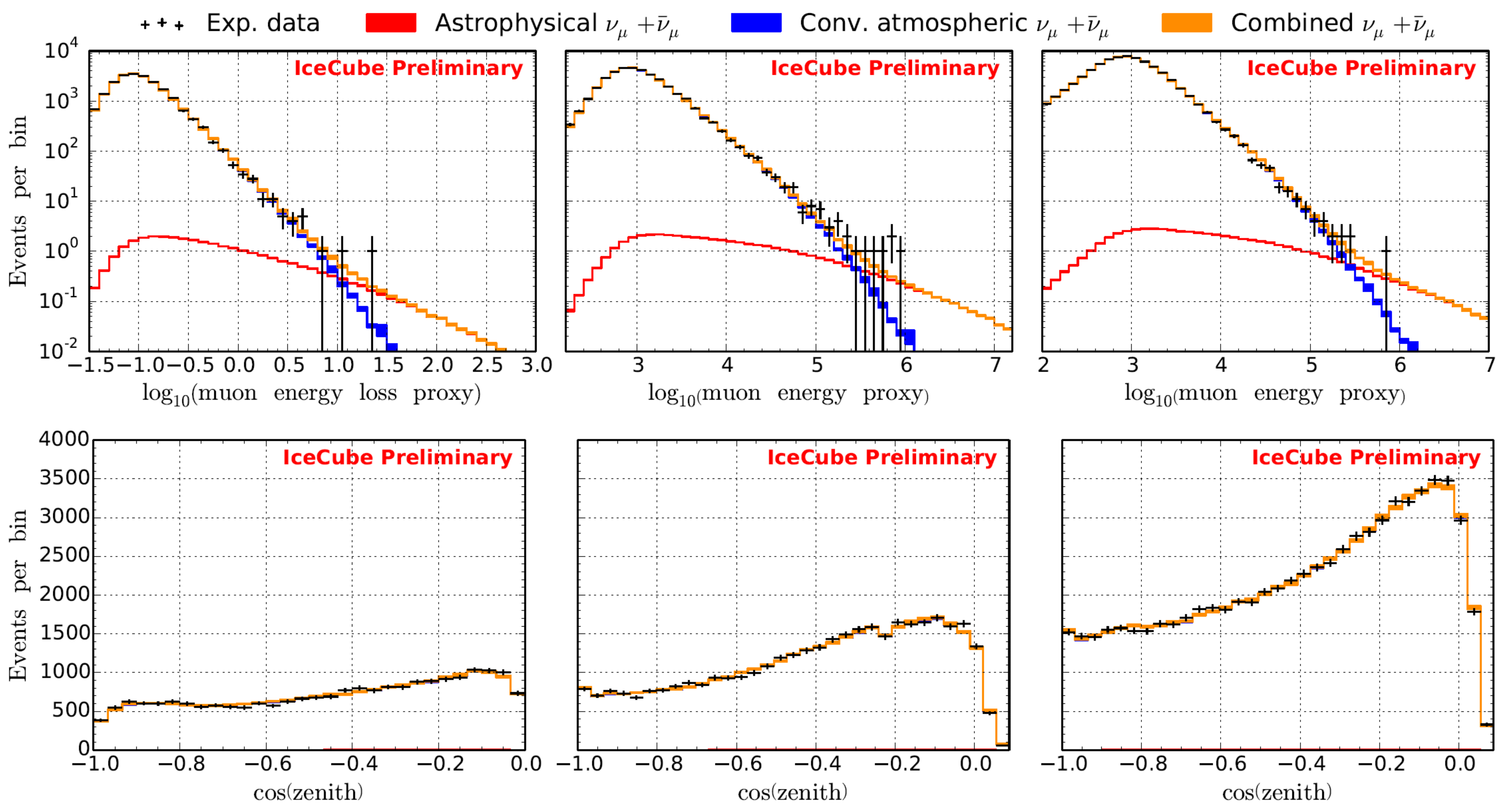} 
\caption{One-dimensional (1D) projections of fit observables weighted by best-fit spectrum for each year.
(Upper row) Reconstructed muon energy proxies. (Bottom row) Reconstructed zenith angle distributions.
(Left to right) 59-, 79-, and 86-string configurations.
}
\label{fig:1079}
\end{figure*}
One of the common methods of rejecting a large amount of downward-going atmospheric muons is to use the Earth as a background shield,
which is achieved by selecting upward-going trajectories of the events for which atmospheric muons would eventually be
absorbed. A high-purity muon neutrino sample can be produced by reconstructing the particle direction and selecting only well-identified upward- and horizontally-going events, which results in accurate
reconstruction of the particle direction.

Recent results from IceCube have been used to update the previous upward-going muon track analysis using three years' worth of samples, which were obtained with different detector configurations from May 2009 to May 2012 \cite{ICRC2015_1079}.
These samples were previously analyzed separately, as one- and two-year datasets.
The first hint at the 1.8$\sigma$ level was found using a one-year IceCube dataset, which was obtained using the partial detector 
in the 59-string configuration \cite{ic59numu}.
The second analysis, performed using two years' worth of data from IceCube and the 79- and 86-string configurations, excluded the  
atmospheric origin of the observed flux only, with a probability of 3.7$\sigma$ \cite{numu}.
In the current analysis, the analysis scheme was unified and a combined data-sample analysis was performed, with the prospect of 
including data from three more years, obtained with a fully configured detector.
After obtaining a high-purity neutrino sample by selecting an upward-going track sample, a possible neutrino
contribution beyond the atmospheric neutrino components was estimated based on multi-dimensional fits over muon energy proxies and zenith angle distributions for
the data sample, with models of different spectral indexes, zenith angle distributions, and flavors.
The best-fit astrophysical flux for the first three-year sample was $\phi(E_\nu) = (0.66^{+0.40}_{-0.30})\dot 10^{-18}{\rm GeV}^{-1}{\rm cm}^{-2}{\rm sr}^{-1}{\rm s}^{-1}(E_\nu/100{\rm TeV})^{-(1.91\pm0.20)}$.
The corresponding significance for rejection of a flux-model hypothesis composed of atmospheric neutrinos only is 4.3$\sigma$.
Figure~\ref{fig:1079} shows the energy proxies and zenith angle distributions of the three-year upward-going muon neutrino
sample with fitted background atmospheric neutrino models and astrophysical
neutrino flux signal models.
An additional three-year full-operation sample is anticipated and it is also reported that an upward-going track with a reconstructed energy deposit of 2.6$\pm$0.3~PeV will be observable in this additional three-year sample \cite{kloppo}.

\subsection{Cascade-event analysis}
\begin{figure*}[!h]
\centering
  \includegraphics[height=2.8in]{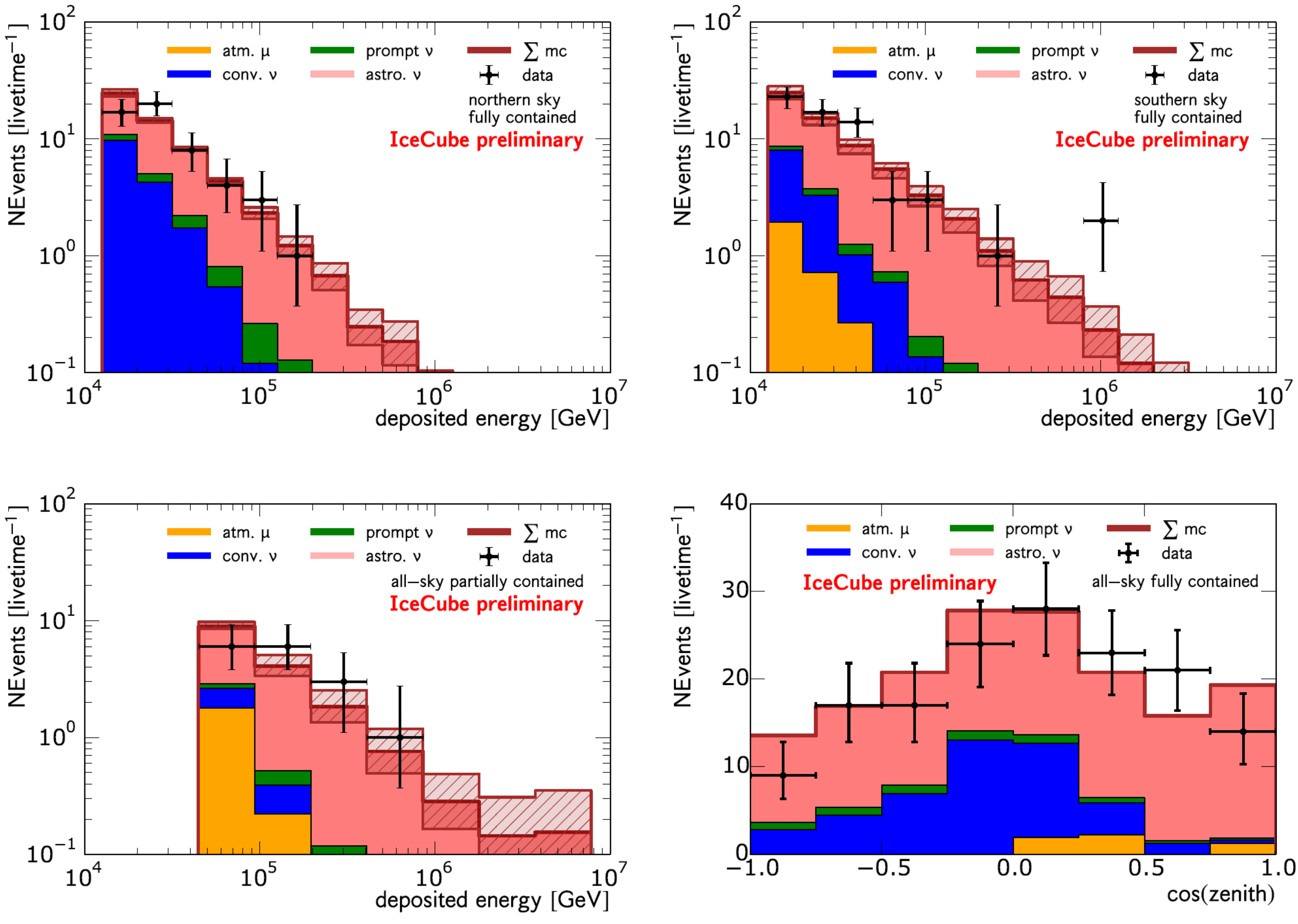} 
\caption{Electromagnetic particle-shower equivalent-energy deposits for fully contained cascades in northern (upper left) and southern (upper right) skies, and for all-sky partially contained cascades (lower left). 
The brown lines indicate the sum of the background and the best-fit astrophysical fluxes in the form of an unbroken power-law model. The hatched regions indicate $1\,\sigma$ uncertainty on the sum of the background and astrophysical fluxes. The zenith angle distributions for fully contained cascades are also shown (lower right). 
}
\label{fig:1109}
\end{figure*}
A second means of reducing the atmospheric muon background is to identify the topological features of
neutrino-induced interactions within the IceCube detector volume.
Cascade-event samples are formed by selecting events with point-like light emissions compared to IceCube's optical sensor separation of 17 and 125 m in the vertical and horizontal directions, respectively.
This cascade channel has full sky coverage and is sensitive to a neutral current of all flavors, along with CC interactions of electron neutrinos and 
tau neutrinos with energy of less than a few PeV; thus, it is complementary to the upward-going muon channel, which is sensitive to the muon neutrino flavor and neutrinos
from the northern sky only.
The track events feature a good pointing resolution with larger energy resolution for the parent neutrino energies, whereas cascade events have good energy resolution with larger directional resolution for the reconstructed incoming neutrino directions.
The other primary difference is that, because electron neutrinos from conventional atmospheric neutrinos are significantly smaller than muon neutrinos, the transition energy from atmospheric to astrophysical neutrinos is largely reduced in the cascade sample compared to that of muon neutrinos.
This allows us to extract high-purity astrophysical neutrino samples over a larger energy range.

The cascade channel analysis has been updated using the two-year data sample obtained from May 2010 to May 2012 at IceCube, which has an effective 
lifetime of 641~days \cite{ICRC2015_1109}.
Along with the use of significantly enhanced statistics, the present analysis has an improved sensitivity to cascade events 
compared to the earlier analyses. This is achieved by selecting cascade events that are partially contained in the detector volume in addition to
fully contained cascades. 
Therefore, the effective detection volume is increased by a factor of two in the energy region above 100~TeV.

The two-year fully and partially contained cascade sample is composed of 172 events, including 152 fully contained cascade events with 
a threshold of 10~TeV and 20 partially contained events above 35~TeV.
The reconstructed energy deposits and zenith angle distributions 
obtained for the two-year cascade sample, with fitted background atmospheric neutrino models and astrophysical
neutrino flux signal models, are shown in the panels of Fig.~\ref{fig:1109}. It was found that this cascade sample is dominated by astrophysical neutrinos ($\sim65\%$) over the entire energy range under study. 
The best-fit astrophysical flux on the sample is 
$\phi(E_\nu) = (2.3^{+0.7}_{-0.6})\dot 10^{-18}{\rm GeV}^{-1}{\rm cm}^{-2}{\rm sr}^{-1}{\rm s}^{-1}(E_\nu/100{\rm TeV})^{-(2.67^{+0.12}_{-0.13})}$, where we observe a softer best-fit spectral index compared to the upward-going muon channel.

\subsection{High-energy starting event analysis}
\begin{figure*}[!h]
\centering
    \includegraphics[height=1.6in]{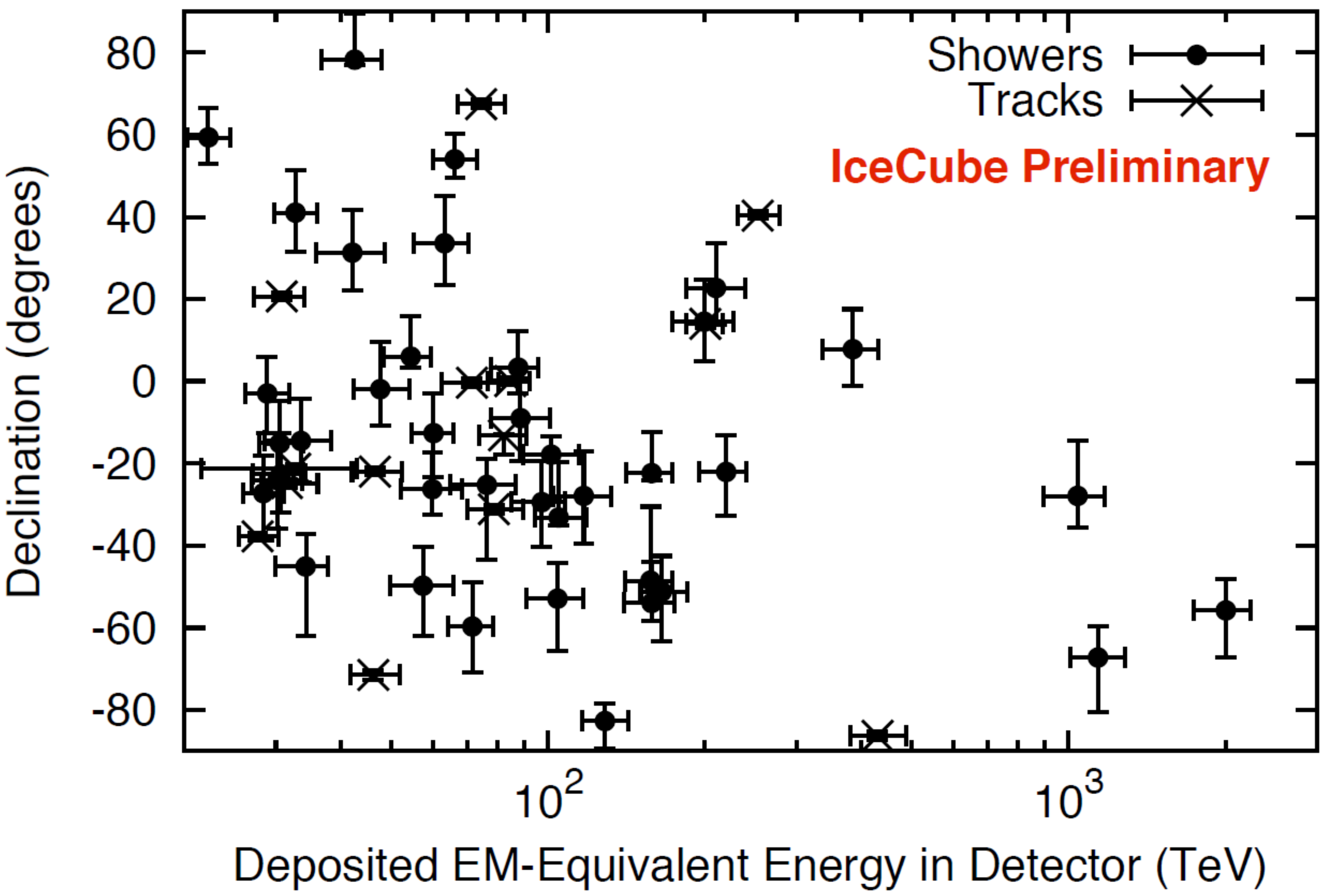}
    \includegraphics[height=1.6in]{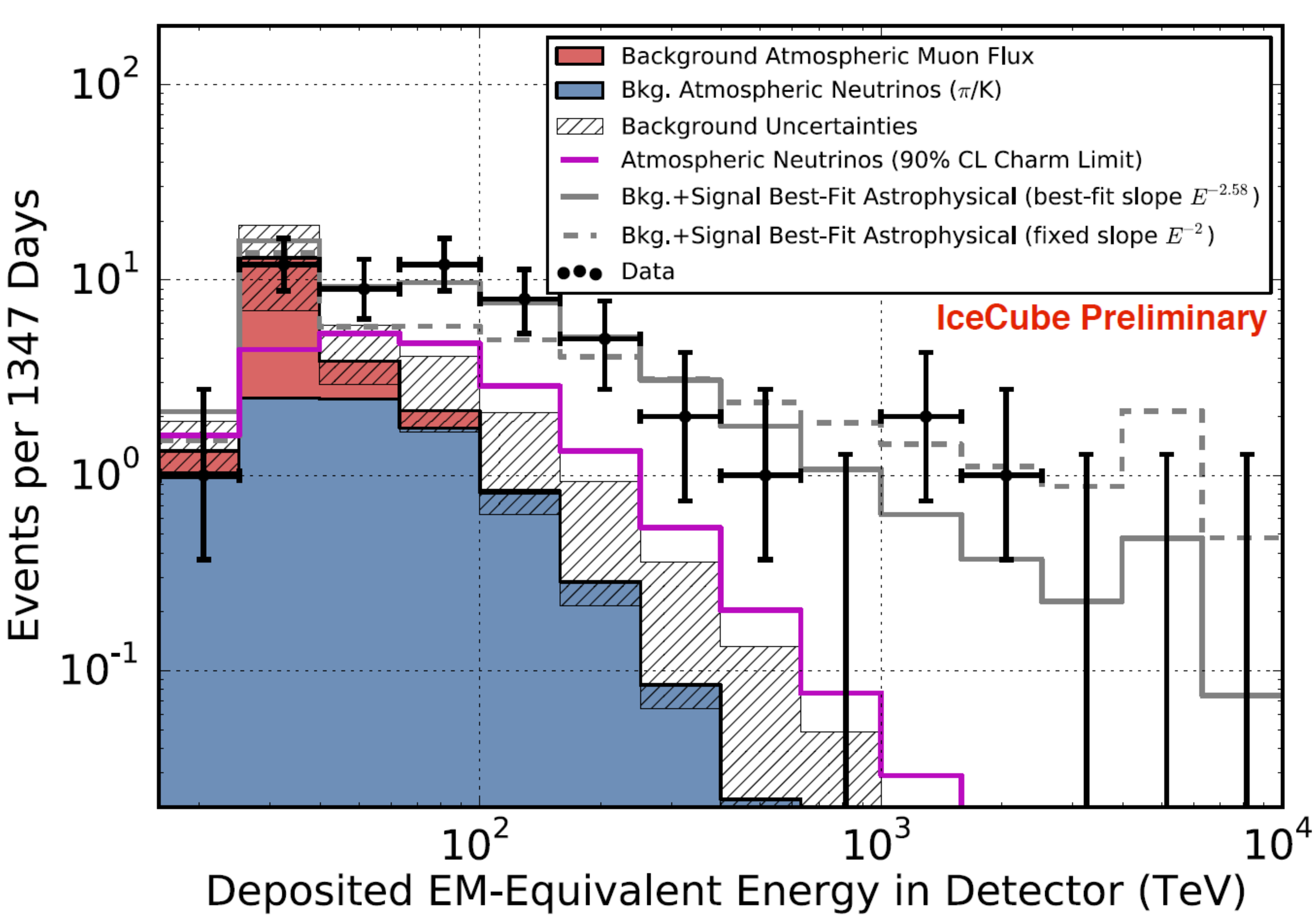}\\
    \includegraphics[height=1.6in]{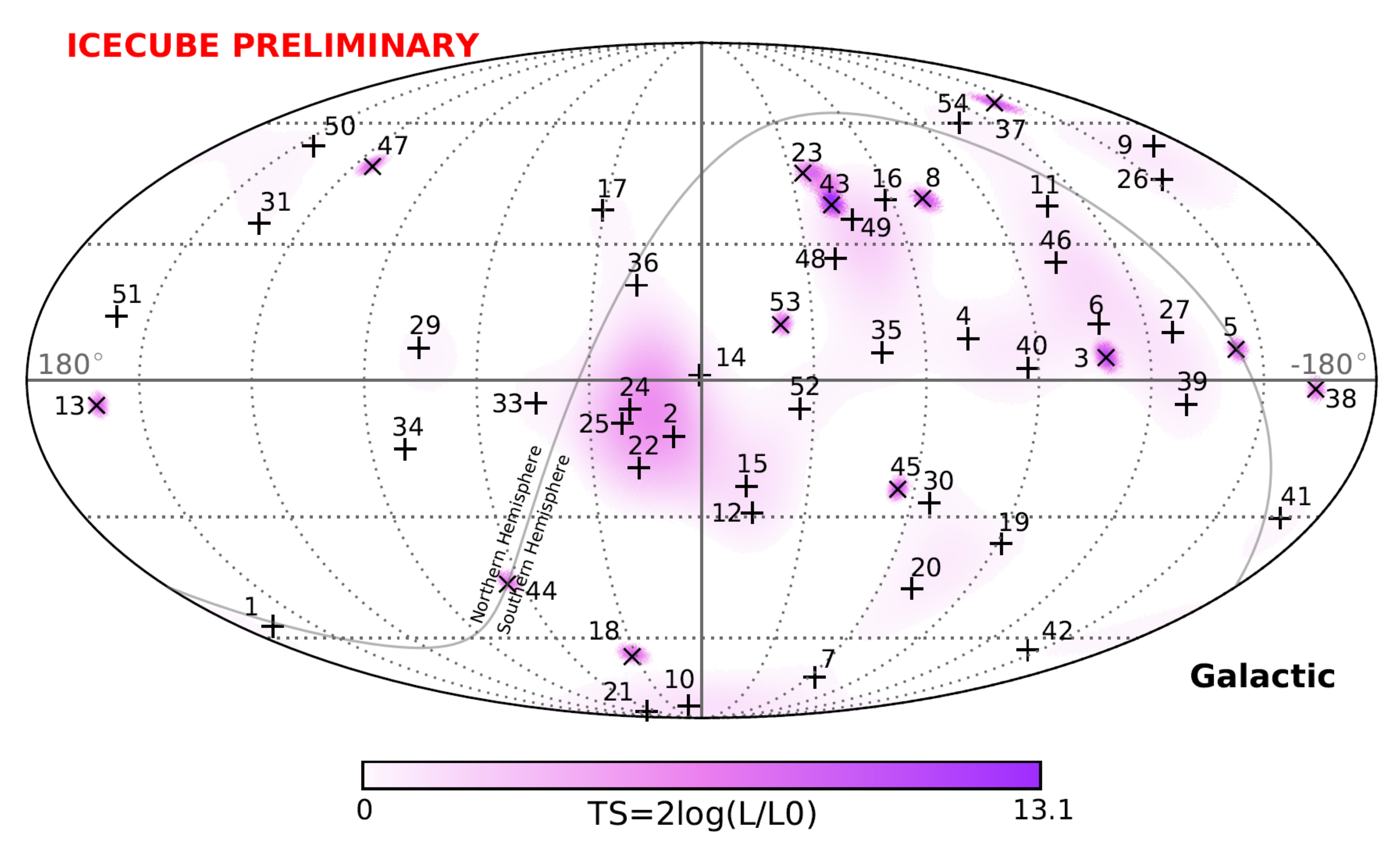}	
\caption{(Upper left)  Distributions of one-year partial configuration and three-year data from the full detector configuration as functions of 
	 reconstructed arrival direction and deposited 
	 energies (electromagnetic particle-shower equivalent). The crosses and filled circles indicate track- and shower-like events, respectively. The error bars show 68\% confidence 
	 intervals including statistical and systematic errors. 
	 (Upper right) Reconstructed deposited
	 energy compared to the background model expectations. The gray lines indicate the sum of the background and the best-fit astrophysical fluxes in the
	 form of an unbroken power-law model. 
	 The dashed and solid lines show a fixed-index spectrum of $E^{-2}$ and a best-fit spectral index of $E^{-2.58}$, respectively.
	 (Lower) Event arrival directions in galactic coordinates. The $+$ and $\times$ symbols indicate track- and shower-like events, respectively. 
	 The colors show the test statistics (TS) for the point-source clustering test at each location. No significant clustering was found.
}
\label{fig:1081}
\end{figure*}
Another relatively simple and, thus, robust method for the selection of neutrino-induced events inside the IceCube instrumented
volume, with the rejection of atmospheric muon events, uses the IceCube exterior as an atmospheric muon veto detector.
The veto method is also sensitive to full-sky astrophysical neutrinos and can reduce the background atmospheric 
neutrino-induced events in the southern sky using a self-veto effect \cite{selfveto}.
For a high-energy starting event (HESE) analysis, high-energy neutrino interactions inside the detector are selected by requiring that the early photons in each event
are well contained. If a sufficient number of early photons is observed in the outer layer (veto region), they are rejected. In addition, a
6,000-p.e. (photo-electron threshold) is imposed in order to select high-energy events of $\sim$ 30 TeV. This simple technique facilitates sampling with more than
50\% purity, with the purity increasing at higher energies.

The results of the HESE sample analysis using four years' worth of data were presented at ICRC 2015 \cite{ICRC2015_1081}. This corresponds to an additional year of data compared to the previous publication. Overall, 54 events were observed with the expected atmospheric background of 12.6 $\pm$ 5.1 muon events and 9.0$^{+8.0}_{-2.2}$ atmospheric neutrino events.
The variable spectral index fit yielded a
best-fit spectral index of
2.58 $\pm$ 0.25, which is softer than the corresponding best-fit index of
2.30 $\pm$ 0.3
obtained for the previous three-year dataset \cite{hese3year}. However, note that the new fit is compatible with the three-year result within the acceptable error range.
As regards the event distribution, the lack of PeV-energy events in the fourth year dataset, in combination with
the comparatively high yield of events in that year, resulted in a steeper spectral fit, as shown in Fig.~\ref{fig:1081}.
The lower panel of Fig.~\ref{fig:1081} shows the arrival direction distribution of the reconstructed HESE sample, which did not yield significant evidence of clustering with p-values of 44\% and 58\%
for the shower-only and all-events tests, respectively. A galactic plane clustering test using a fixed width of 2.5$^\circ$
around the plane yielded a p-value of 7\%, and a p-value of 2.5\% was obtained using a variable-width scan. All the above p-values were corrected for
trials.

\section{Point-source and multi-messenger analysis}

Although precision measurements of diffuse cosmic neutrino spectra certainly provide information on the 
properties of the UHECR production mechanism,
the ultimate goal of neutrino astronomy is to observe neutrino-emitting point sources in the universe.
As 
neutrinos with energies of up to the exavolt (EeV; 1~EeV
= $10^{18}$ eV) range are expected to be produced in cosmic-ray 
acceleration processes at the source site via cosmic-ray interactions with surrounding photons or matter,
high-energy cosmic neutrinos are expected to constitute one of most direct tools for identification of the origin of
high-energy cosmic rays. 
As is well known, a neutrino
traverses a cosmological distance without being deflected
by a cosmic magnetic field or absorbed by a photon field; thus,
a straightforward interpretation of the neutrino arrival direction as indicating the 
cosmic-ray origin is possible.
Observations of neutrino-source candidates with other probes, namely, gamma and cosmic rays, 
may play a key role in the discovery of neutrino objects, as these rays are related by well-known interactions.
If the diffuse neutrino flux component observed by IceCube is from a transient object, the multi-messenger approach 
will be even more relevant.

Many improvements have been made to the methods used in the search for neutrino point sources, although positive detection 
of these sources is yet to be achieved; these refinements were also reported at the ICRC 2015 conference.
Further improvement can be achieved by analyzing more data and improving the sample angular resolution. The point-source search at IceCube, which includes six years' worth of data samples, the similar search at the Astronomy with a Neutrino Telescope and Abyss Environmental Research Project (ANTARES) detector, which uses samples including both track- and cascade-like events to enhance the statistics, and the joint point-source analyses of the IceCube and ANTARES samples are moving in this direction.
Another attempt at point-source analysis utilizes the selected source candidate list.
At the ICRC 2015 conference, an update on an interesting study on neutrino emission in the Fermi bubble region was provided by researchers at the ANTARES detector. 
%
\subsection{Six-year IceCube point-source search}
\begin{figure}[!h]
\centering
  \includegraphics[height=2.in]{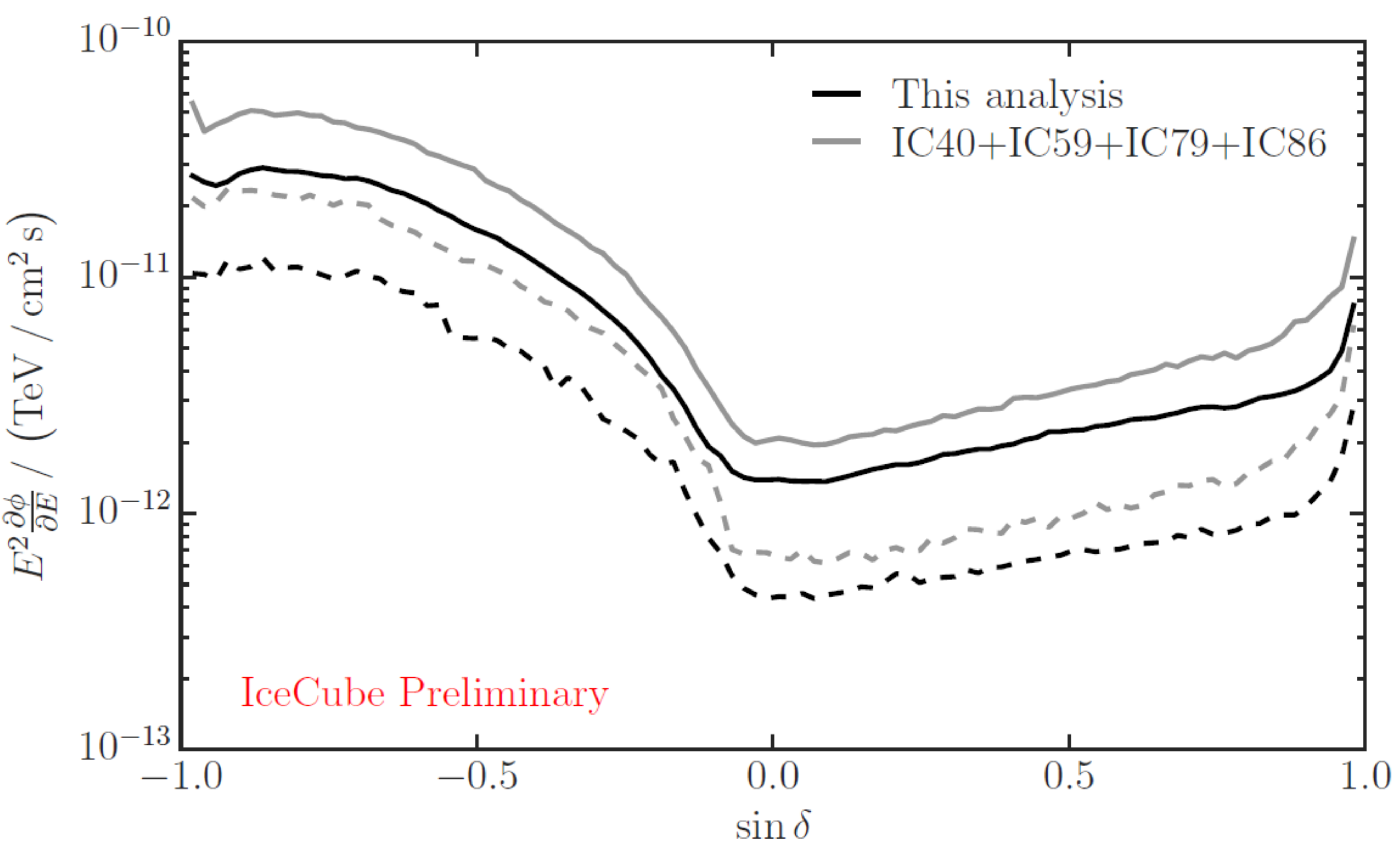} 
\caption{Six-year point-source
IceCube sensitivity (dashed) and discovery potential (solid)
             for steady point-like sources for six years' worth of accumulated data. This
             analysis (black) adds two years to the previous
             analysis \cite{4year_ps} (gray).
             The black crosses indicate $90\%$ upper limits for
             selected source candidates at their declinations.
}
\label{fig:1047}
\end{figure}
At IceCube, a point-source search based on six years' worth of sample data (from April 2008 to May 2014) has been performed \cite{ICRC2015_1047}. 
This corresponds to an inclusion of two years' worth of additional samples compared to the previous analysis.
This yields a sample of approximately
$600\,000$ events.
The sensitivity of the IceCube detector is
optimal for neutrino point-like source emission in the
TeV--PeV and PeV--EeV ranges in the northern and southern skies, respectively.
IceCube results a current highest neutrino point-like source
sensitivity at energies above 100\,TeV.

For the first time, the northern sky sensitivity has moved below
$10^{-12}\,\mathrm{TeV\,cm^{-2}\,s^{-1}}$ over the majority of the IceCube declination range, as shown in Fig.~\ref{fig:1047}.
In the analysis of the six-year upward-going track data sample integrated over a lifetime of
$2\,063$ days,
an unbiased all-sky scan and a search for
neutrino emission from a limited list of known non-thermal objects were performed.
In the analyzed data, no clustering significantly above background expectation was
found. Further, the most interesting points in the all-sky search and the
gamma-ray source list were compatible with background expectations.

\subsection{Combined muon track and cascade ANTARES point-source search}
\begin{figure*}[!h]
\centering
  \includegraphics[height=2.in]{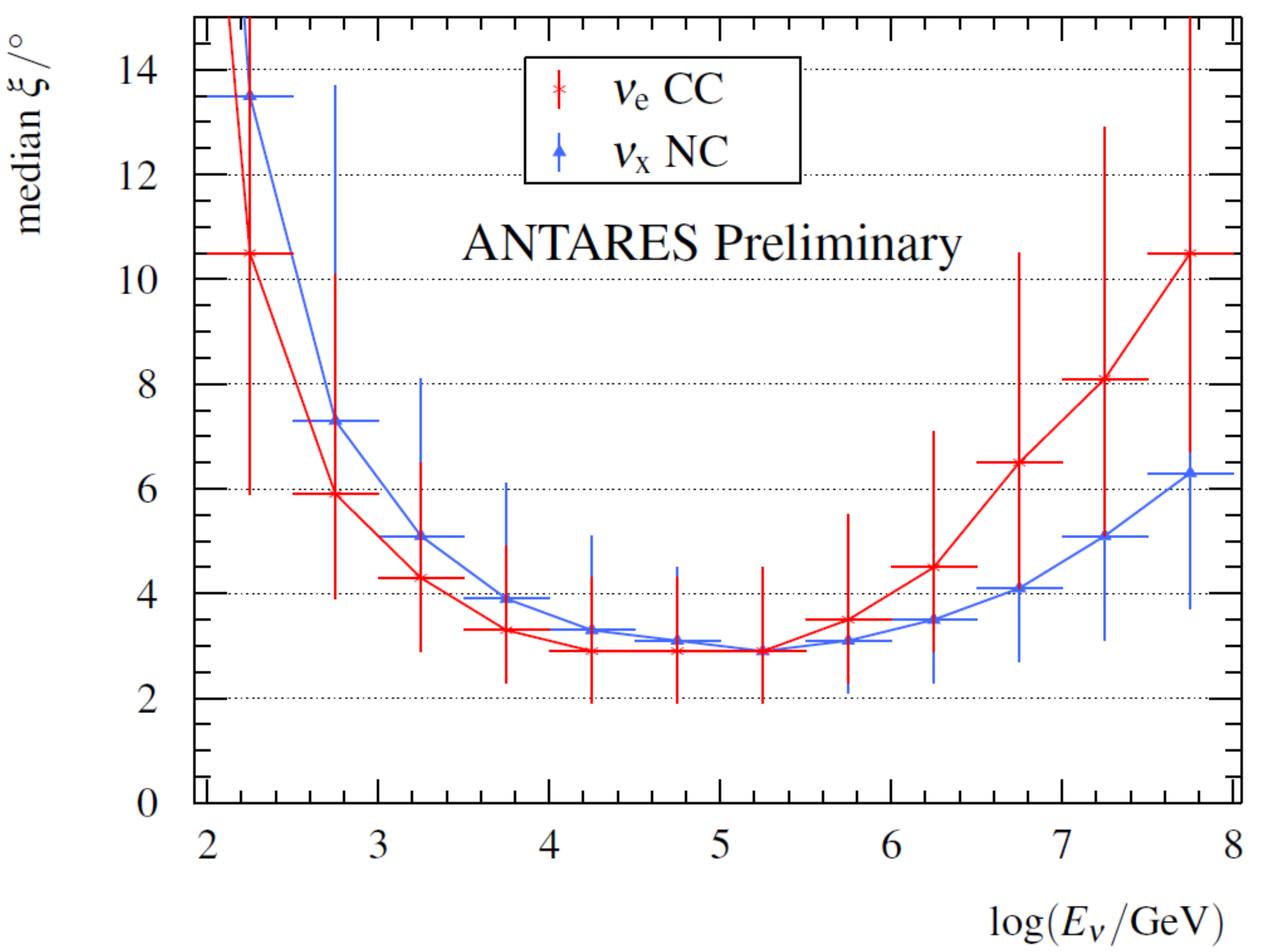} 
  \includegraphics[height=2.in]{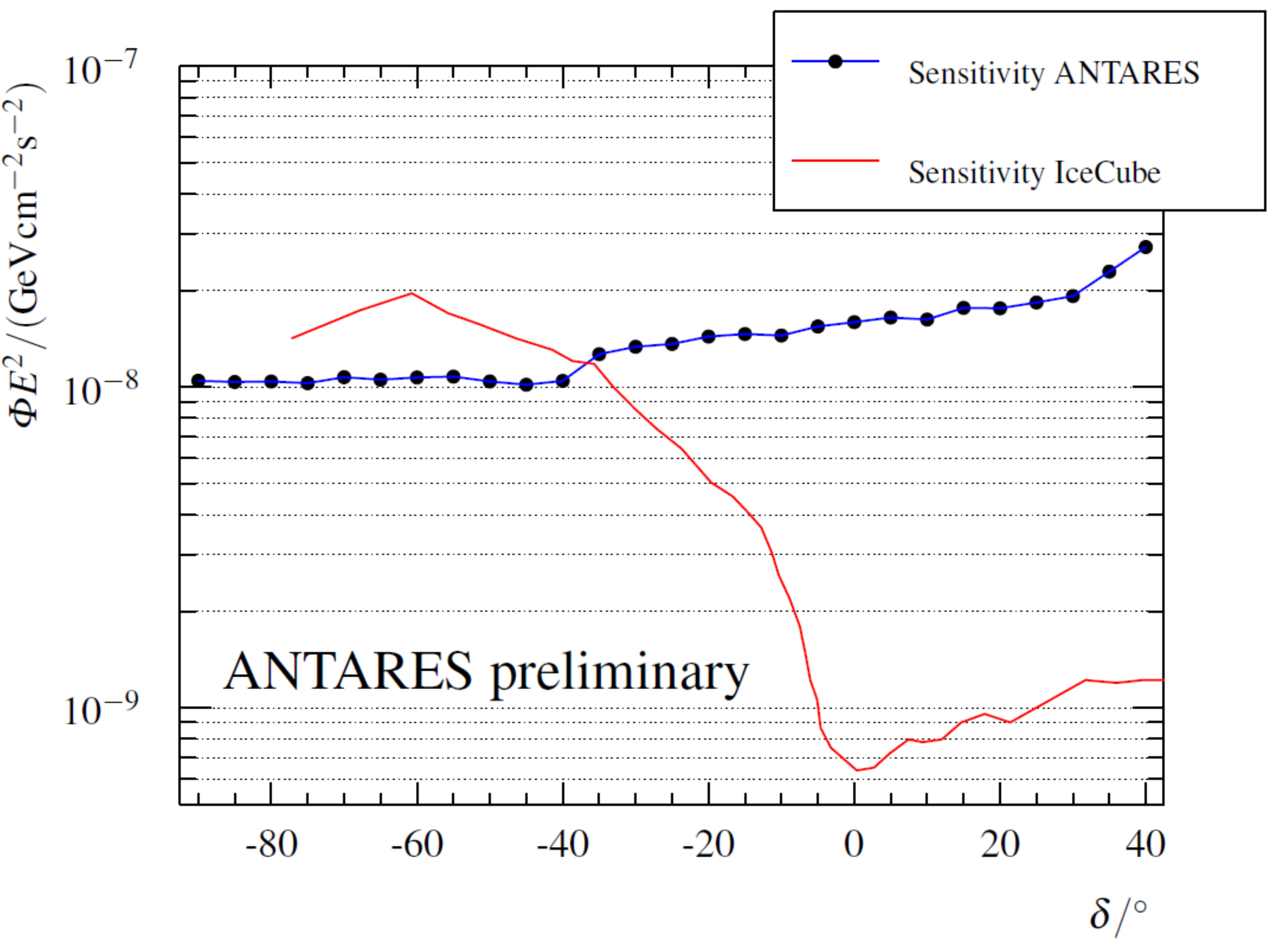} 
\caption{(Left) Shower directional reconstruction performance, where the 
red and blue lines are for electromagnetic and hadronic showers, respectively.
It is apparent that an angular median resolution of $3^{\circ}$ is expected 
in the 10--100-TeV energy range, which is the most important 
energy region as regards neutrino point-source searches.
(Right) ANTARES sensitivity (blue) for steady point-like sources using 
1622 days' worth of ANTARES samples and IceCube sensitivity for four years' worth of IceCube samples with a 1373-day lifetime.}
\label{fig:1078}
\end{figure*}
The ANTARES telescope is a deep-sea neutrino detector located in the Mediterranean Sea, which has been acquiring 
data in its final configuration since 2008.
Although the instrumented volume of ANTARES is a factor of approximately 100 smaller than IceCube,  
this detector, being located in the northern hemisphere, is more sensitive to many galactic sources in the energy region below 100~TeV. 
Point-source searches including cascade-like events have been performed at ANTARES,
in addition to the sampling of standardized upward-going track events.
By exploiting the fact that this detector is located underwater, good pointing resolution 
for particle shower events has been established at ANTARES. This allows all-flavor neutrino point-source searches to be performed.
The angular median resolution in the 10--100~TeV energy range is estimated to be $\sim3^{\circ}$, and
the corresponding median resolution for the track events is $\sim0.4^{\circ}$ \cite{ICRC2015_1078, antares_ps}.

The analyzed ANTARES data sample was taken in the period from 2007 to 2013, having a lifetime of 1622 days and 
containing 6261 muon track and 156 cascade events. Both the track and cascade samples were composed of 
10\% atmospheric muons and 90\% atmospheric neutrinos.
From Fig.~\ref{fig:1078},
it can be seen that the selected cascade events are expected to increase in signal event rate and sensitivity by 30\% for an $E^{-2}$ signal flux with equal neutrino flavor composition.
Thus, the combined sample exhibits successfully improved sensitivity. The sensitivity for 1622 days of observation corresponds to 
$\sim10^{-8}\,\mathrm{GeV\,cm^{-2}\,s^{-1}}$ over a declination angle of less than $-40^{\circ}$, which slowly increases to 
 $\sim2\times10^{-8}\,\mathrm{GeV\,cm^{-2}\,s^{-1}}$ towards a declination angle of $20^{\circ}$.

\subsection{IceCube and ANTARES combined-sample point-source search}
\begin{figure*}[!h]
\centering
\includegraphics[height=2.in]{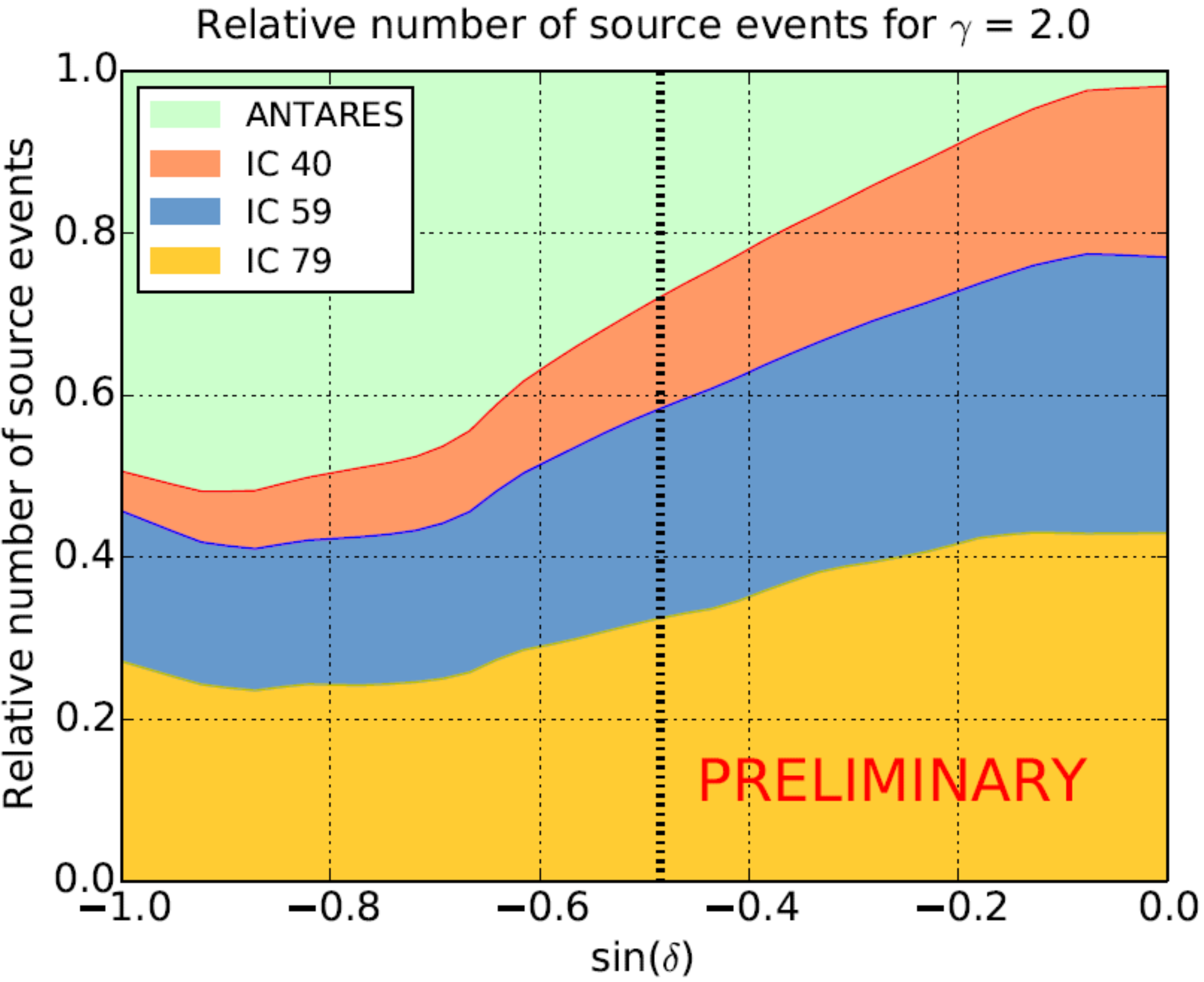}
\includegraphics[height=2.in]{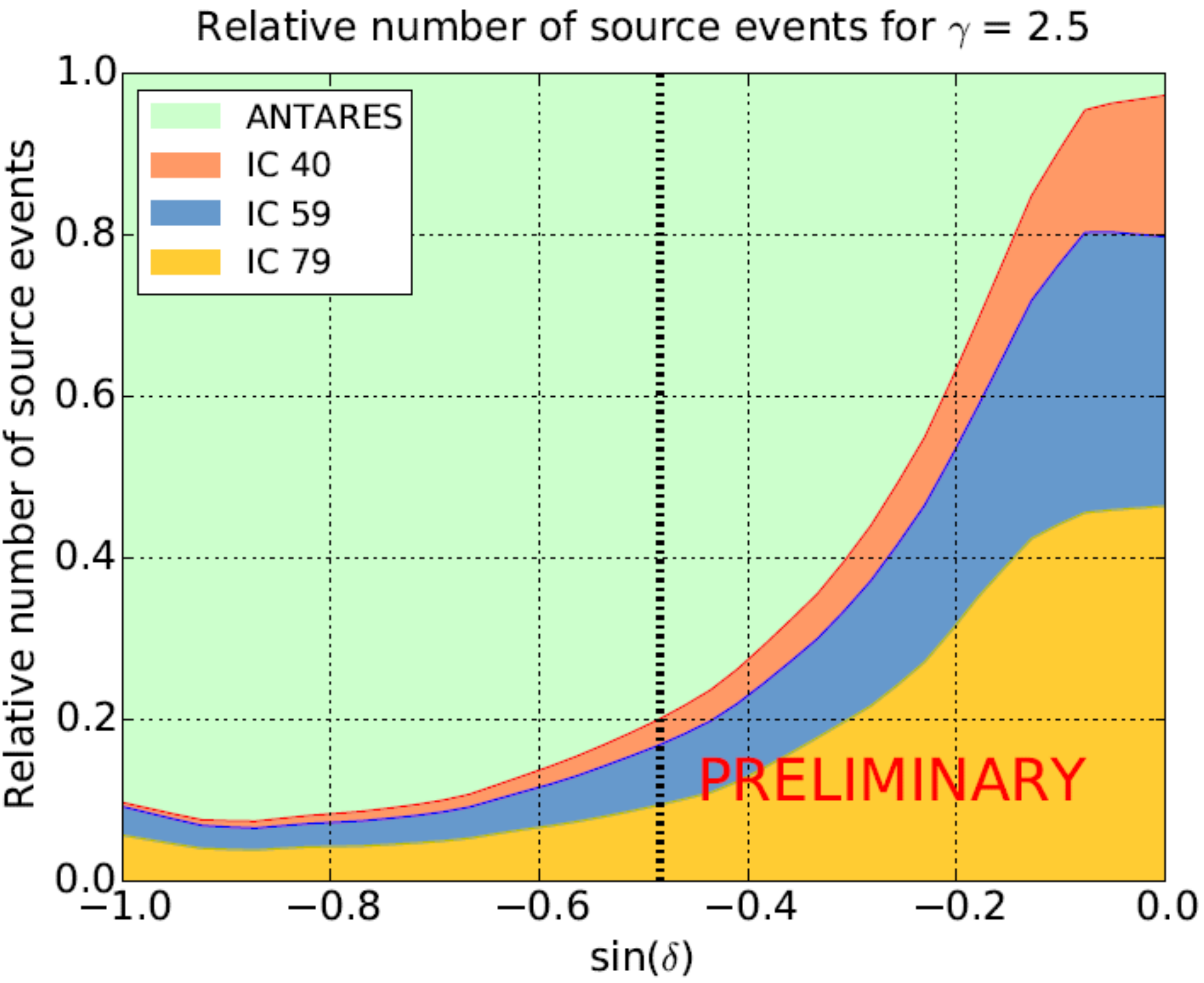}
\includegraphics[height=2.in]{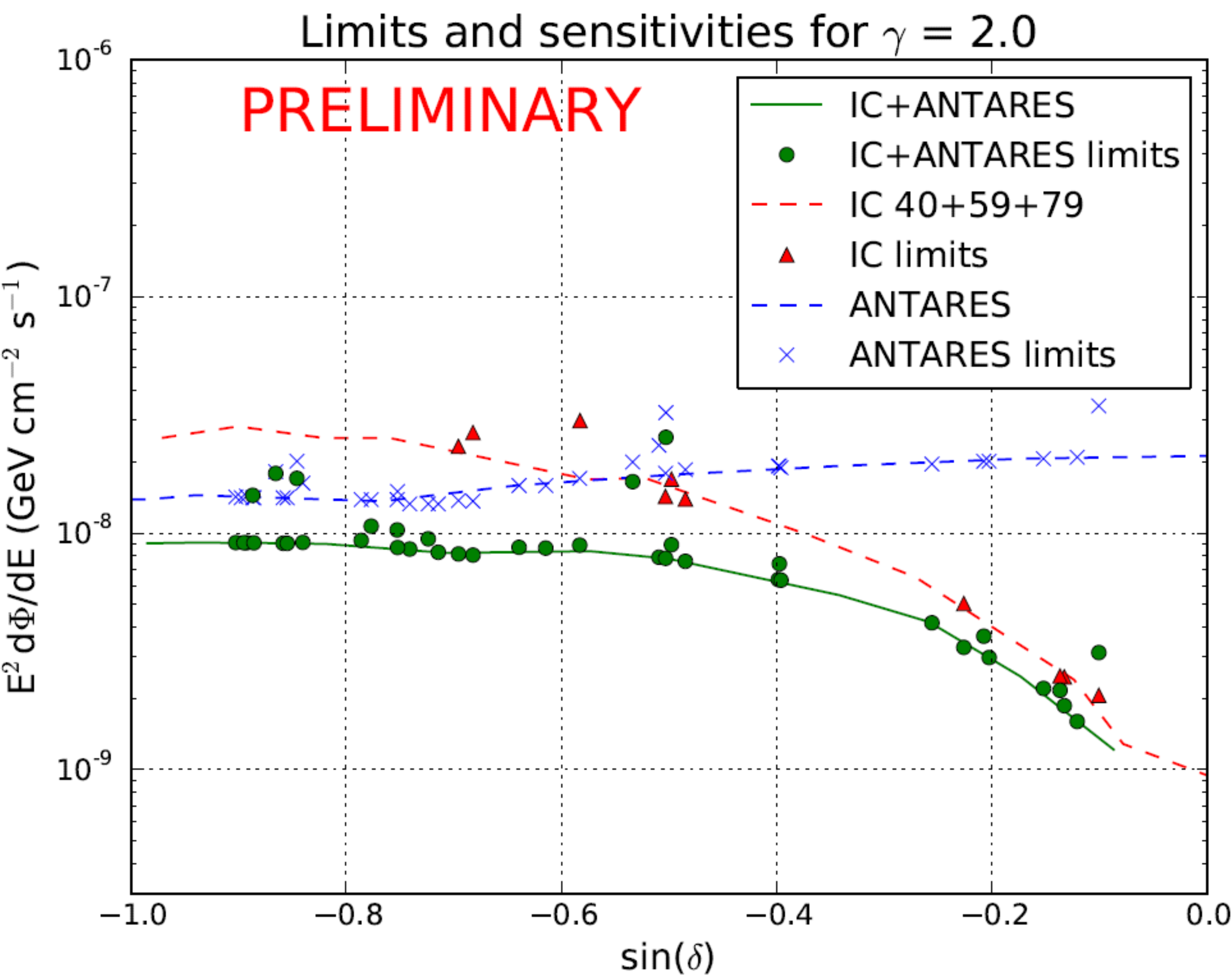}
\includegraphics[height=2.in]{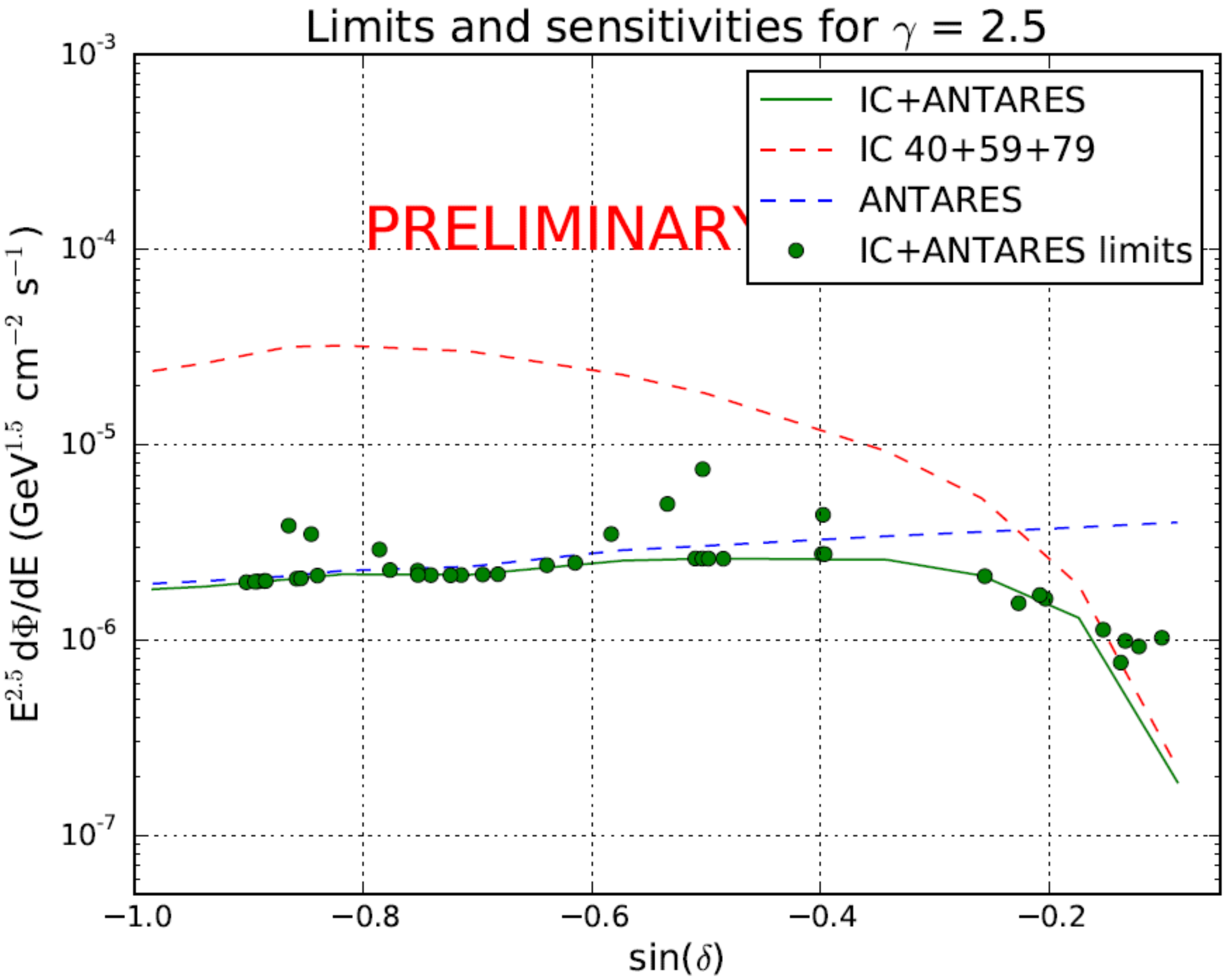}
\caption{(Upper) Relative fractions of ANTARES and IceCube (with different 
detector configurations) sample signal events as functions of source declination. (Lower) 90\% CL sensitivities and limits for neutrino emission from point sources as 
a function of source declination in the sky. $E^{-2}$ (left) and $E^{-2.5}$ (right) power-law spectra are assumed.}
\label{fig:1076}
\end{figure*}
A neutrino point-source search in the southern sky has been conducted using
both ANTARES and IceCube data samples \cite{ICRC2015_1076}.
The IceCube sample constitutes three years' worth of data obtained between April 2008 and May 2011, 
while the
ANTARES sample is from January 2007 to December 2012. 
As noted above, IceCube is a cubic-kilometer-sized detector located at the South Pole.
Although ANTARES is significantly smaller than IceCube, a lower energy threshold is possible (below 100~TeV in particular) as it is located in the northern hemisphere. Thus, the samples are complementary.
For the sample comprising track events from the southern sky, the IceCube data is dominated by atmospheric muons, 
whereas the ANTARES data constitutes atmospheric neutrinos with estimated contamination from mis-reconstructed atmospheric muons of 10\%.
The upper plot in Fig. \ref{fig:1076} shows the relative fraction of signal events 
for unbroken $E^{-2}$ and $E^{-2.5}$ spectra. For $E^{-2}$, there is a significant 
contribution from both the ANTARES and IceCube samples. As regards softer power-law 
spectra (i.e., $E^{-2.5}$), the relative contribution of the lower-energy events, and thus the 
contribution from the ANTARES sample, increases. For power-law spectra harder than $E^{-2.5}$, 
it is shown that combination of the IceCube and ANTARES samples successfully increases
the sensitivity in comparison to those of the individual samples. 
Thus, it has been demonstrated that the combined sample has increased sensitivity to neutrinos emitted from point sources
with different spectral indexes.
Therefore, joint analysis of future datasets will facilitate multi-experimental studies on point-source sensitivity in 
critical overlap regions in the southern sky.

\subsection{ANTARES extended point-source search at Fermi bubble region}
\begin{figure*}[!h]
\centering
  \includegraphics[height=2.1in]{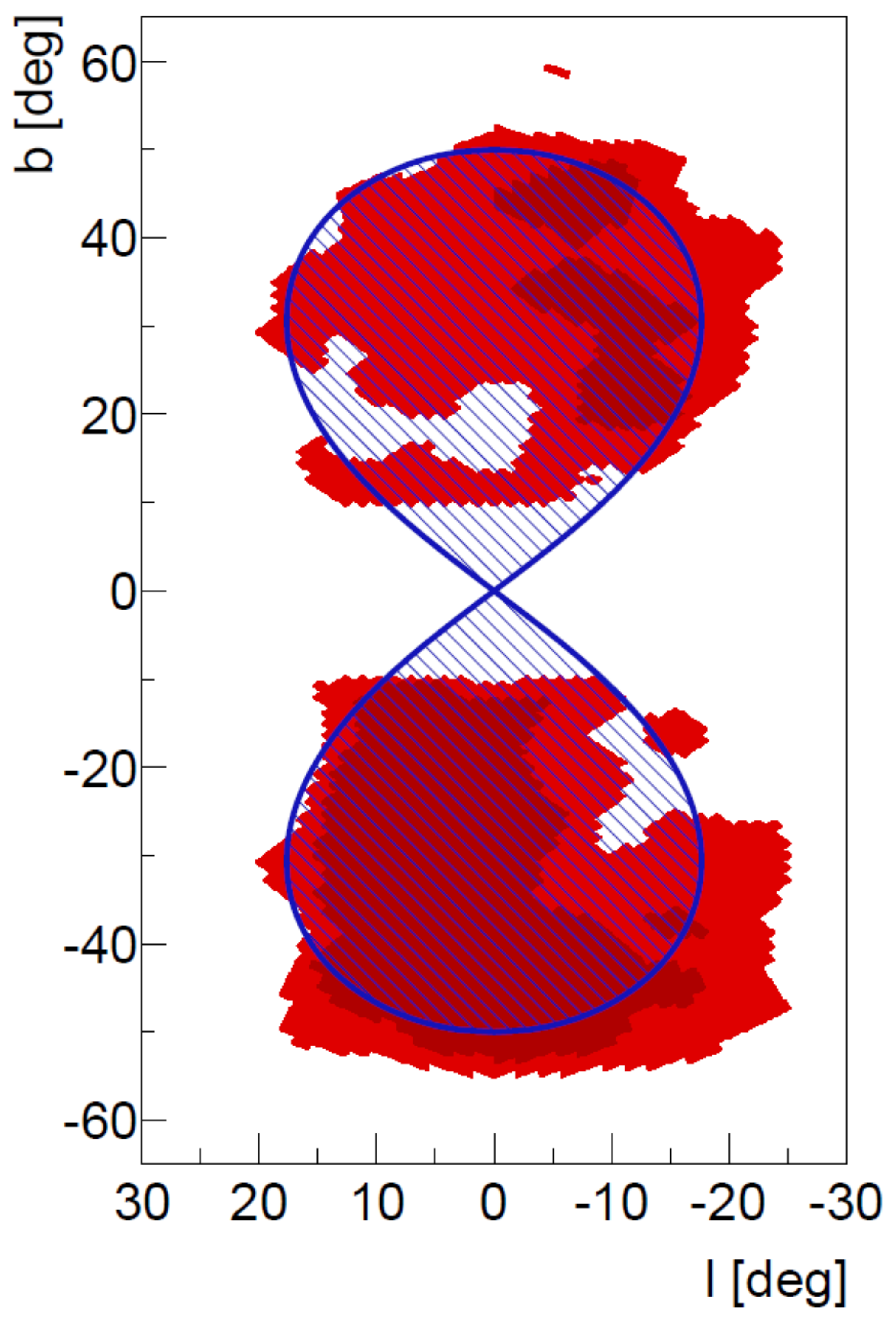} 
  \includegraphics[height=1.8in]{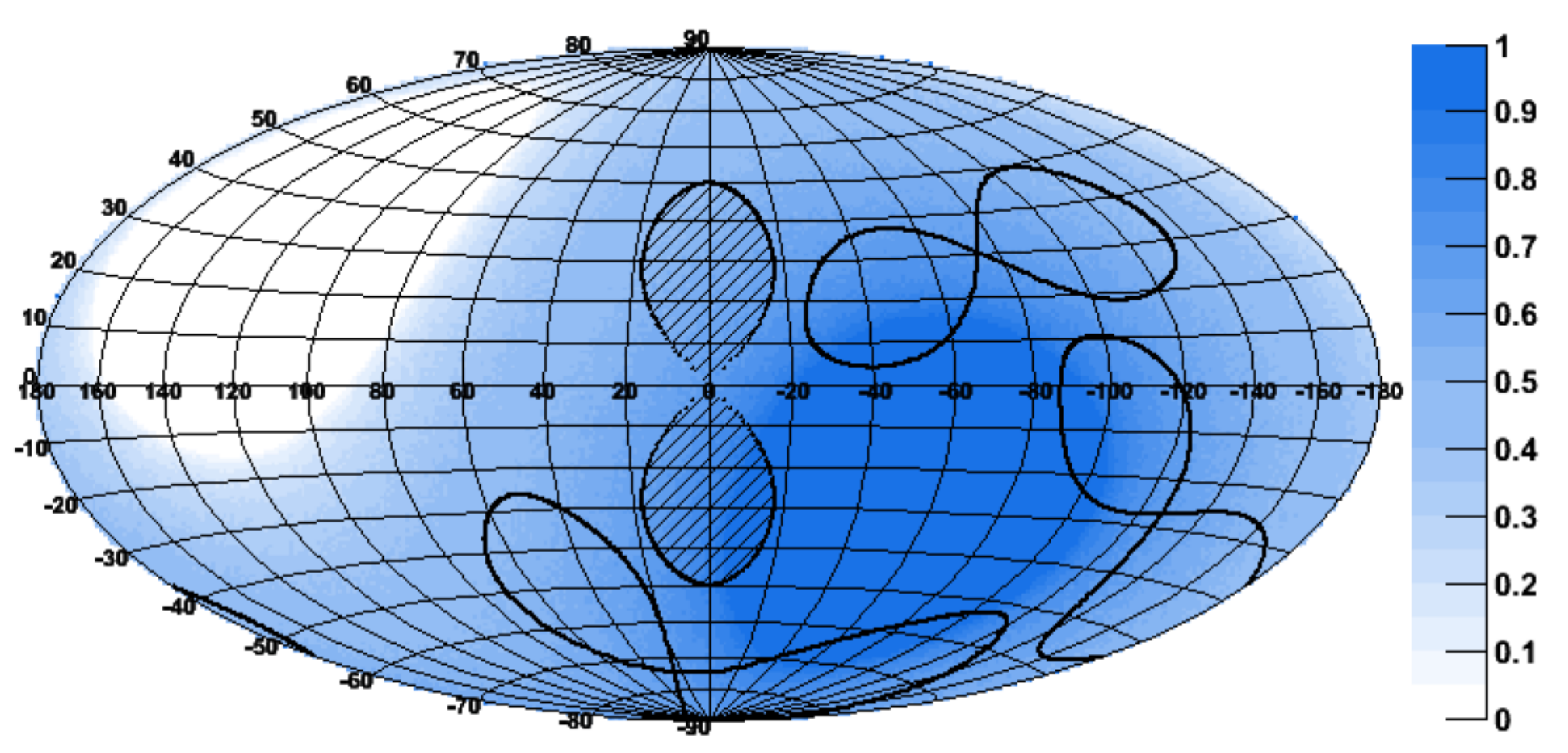} 
  \includegraphics[height=2.in]{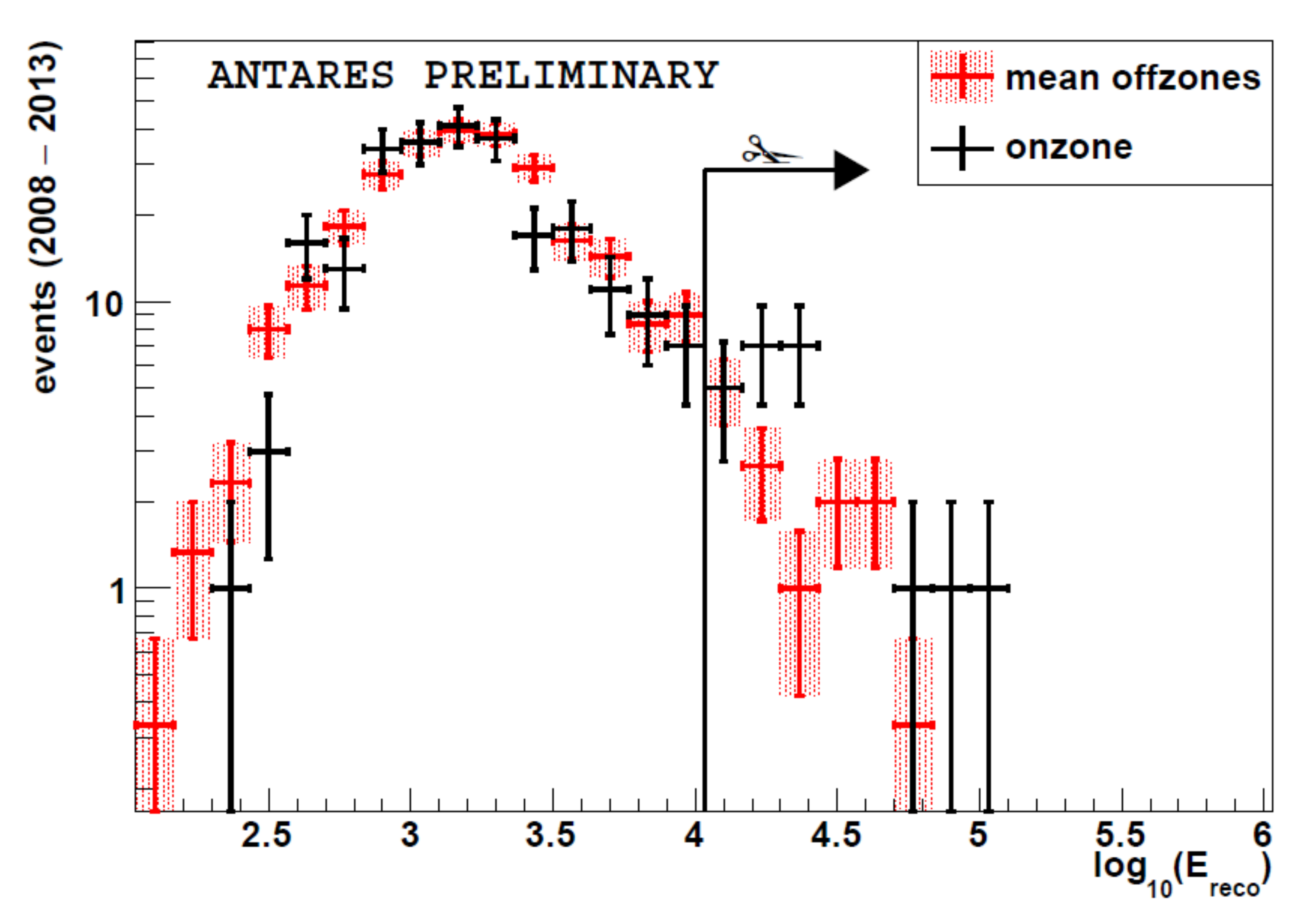} 
\caption{(Upper left) The blue region indicates the geometric shape used in the analysis. (Upper right) The background off-zones and signal on-zone regions are projected in galactic coordinates.
The color scale represents the visibility of the sky at the detector site.
(Lower) Reconstructed energy distribution for six years' worth of ANTARES data. The final level cut is applied at ${\rm log_{10}}({\rm E_{reco}/GeV})\geq 4.03$.}
\label{fig:1059}
\end{figure*}
The Fermi-Large Area Telescope (LAT) experiment discovered two giant lobes of $\gamma$-ray emission extending above and below the Galactic Center, 
which are called Fermi bubbles (FB).
While the origin of the FB remains unknown, this phenomenon could indicate a cosmic-ray acceleration process near the
Galactic Center.
Therefore, the observation of a neutrino signal from the FB region may provide insight into the origin and behavior of the FB.
The ANTARES project has searched for such a signal using a four-year dataset (2008--2011) and setting an upper limit \cite{fb}. The result of an update using two additional years' worth of data was presented at ICRC 2015 \cite{ICRC2015_1059}.
Off-zones with the same visibility in the ANTARES detector were employed in this analysis so as to determine the background.
The upper left panel of Fig.~\ref{fig:1059} shows the shape of the gamma-ray lobes observed with Fermi-LAT and the parameterized area used in the analysis.

In the previous analysis using the four-year sample, there was a statistically insignificant excess of 
1.2$\sigma$ corresponding to 16 events from the signal region. On average, 11 (= 33/3) events 
from off-regions within the same area were observed.
For the six-year ANTARES dataset, the number of events observed in the signal region was 22 over an average number of background region events of 13 (= 39/3), with an increase in the excess to 1.9$\sigma$. Further, an insignificant but consequential observation
of upper fluctuation in the observed on-source region has drawn attention to future updates on this analysis.

\section{Next-generation experiments}

As mentioned above, identifying the high-energy astrophysical 
neutrino flux in the energy region from a few tens of TeV to PeV was a significant success for the field of neutrino astronomy.
However, as has been demonstrated in the previous sections, the potential of neutrino astronomy 
has not yet been maximized because of the currently limited statistics. 
While a Cherenkov detector of 1~km$^3$ is sufficiently large to yield the basic initial properties of 
the neutrino flux, if we wish to confirm the origin of this observed neutrino flux, higher-precision detectors 
are likely to be required.

Therefore, a number of active design studies on next-generation neutrino observatories in the
Antarctic and in the Mediterranean Sea 
are being performed.
These design studies are aimed towards the construction of high-precision detectors of the order of 
10~km$^3$.
Along with their increased size, detectors constructed at the South Pole or in the Mediterranean Sea must
satisfy challenging requirements as regards the specifications necessary for neutrino astronomy.
Experience obtained over recent decades will facilitate the construction of such up-scaled detectors.

The ongoing research on the proposed IceCube-Gen2 \cite{ICRC2015_1146} 
and the Cubic Kilometer Neutrino Telescope-Astroparticle Research with Cosmics in the Abyss (KM3NET-ARCA) high-energy neutrino detectors \cite{ICRC2015_1158}
were reviewed extensively at the ICRC 2015 conference.
Note that the low-energy extensions of these experiments, the Precision IceCube Next Generation Upgrade (PINGU) \cite{ICRC2015_1174} and Oscillation Research with Cosmics in the Abyss (ORCA) \cite{ICRC2015_1140}, are also considered to be 
important components of the full neutrino observatories, 
having the primary physical goal of identifying the neutrino mass hierarchy.
Another 1~km$^3$-scale underwater detector, the Baikal-Gigaton Volume Detector (GVD) \cite{ICRC2015_1165}, is planned for construction 
in Lake Baikal, and the first construction phase has already begun.

The forthcoming major achievement milestone is the discovery of neutrino point sources 
using such next-generation Cherenkov neutrino detectors.
Furthermore, the discovery of any other classes of neutrino fluxes other than those already observed 
may also move neutrino astronomy into the next phase.
In particular, in the high-energy region above 10~PeV,
cosmogenic neutrinos are expected to be produced in interactions between the highest-energy cosmic rays and
the cosmic microwave background through the Greisen-Zatsepin-Kuzmin (GZK) process.
If observed, these neutrinos may be the most direct evidence of the highest-energy cosmic ray sources.
In addition, other on-source astrophysical neutrino fluxes above a few PeV may exist, with 
corresponding cosmic-ray energies slightly below the ankle of the cosmic-ray spectrum and above.

One of the most promising and cost-effective means of detecting these possibly low-level but extremely high-energy neutrino
fluxes is considered to be neutrino signal detection in ice using radio antennas. This is because 
the Cherenkov emission from a neutrino-induced large cascade induces detectable 
radio emission that is coherently enhanced via the Askaryan effect.
In contrast to optical light, of which the absorption length in ice is {\it O}(100)~m, 
radio waves travel for a distance of more than 500~m (or even more than 1 km) in ice, depending on the temperature.
Thus, these signals are measurable using very sparse and cost-effective radio antennas.

\subsection{IceCube-Gen2}
\begin{figure*}[!h]
\centering
  \includegraphics[height=2.3in]{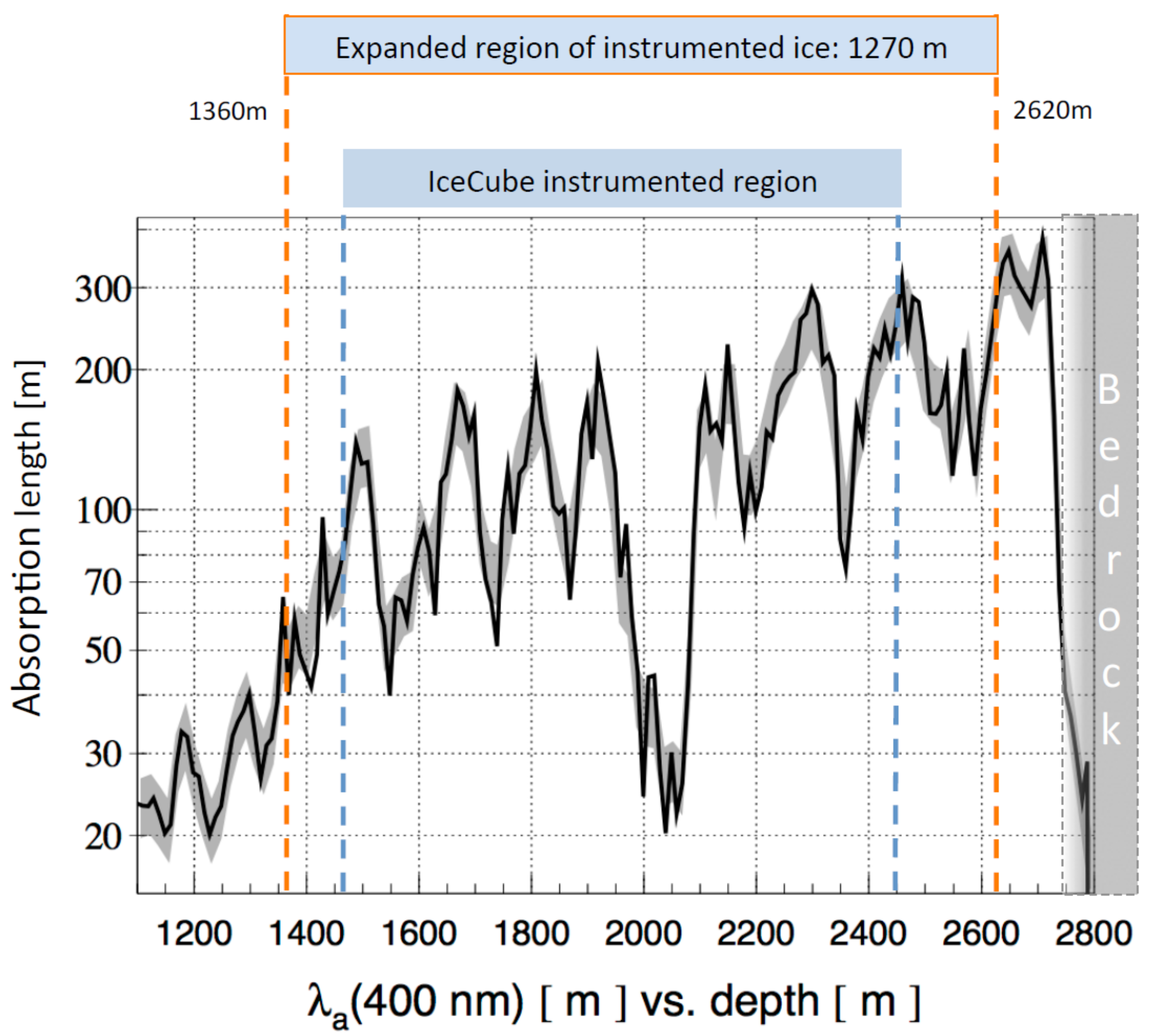} 
  \includegraphics[height=2.3in]{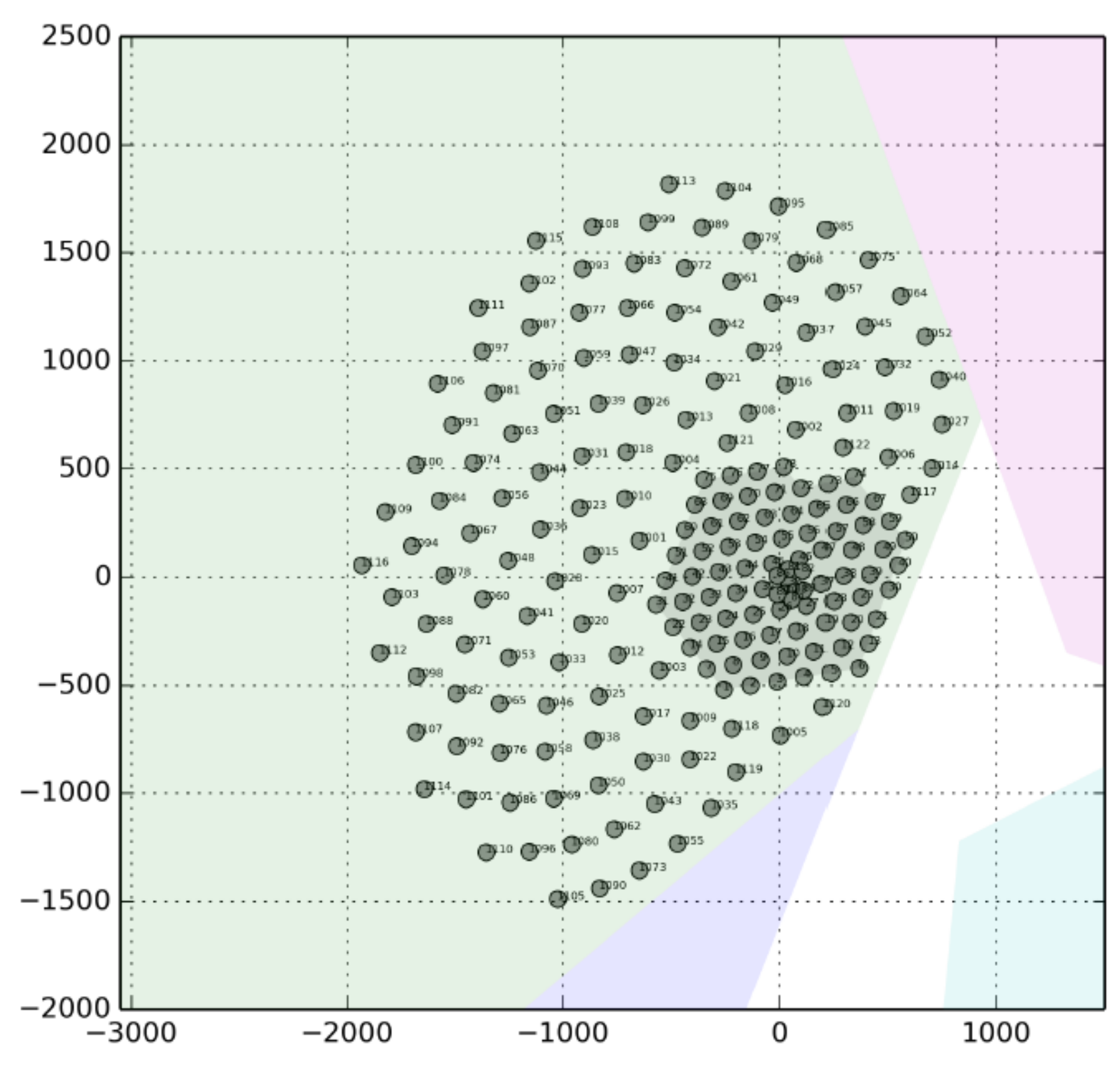} \\
    \includegraphics[height=2.2in]{c741_ICRC2015_1146_Fig4.pdf} 
\caption{Volume extension of IceCube-Gen2 compared with current IceCube detector in vertical and horizontal directions. 
(Upper left) Absorption length in glacial ice versus depth. The current instrumented depth range used in IceCube and the extended
string length, with approximately 260 m added to each string, are
indicated.
(Upper right) Sample benchmark detector string layout under study, with
120 strings added to IceCube. This example shows a uniform 
string spacing of
240 m and the instrumented volumes are 6.0 km$^3$.
Other configuration options are under consideration and are being optimized for physical outputs.
(Lower) Expected 26-PeV muon-neutrino-induced muon track event for Gen2.
}
\label{fig:1146}
\end{figure*}
IceCube-Gen2 is a Cherenkov neutrino detector that will be constructed at the South Pole near the existing IceCube detector \cite{ICRC2015_1146}. The former will have 
120 additional holes occupied by optical modules in the lower 1270 m for a detector position 1360~m below the surface.
The upper left plot in Fig.~\ref{fig:1146} indicates the ice absorption length as function of depth and the instrumentation
depth of the IceCube-Gen2 detector.
Gen2 will be used to measure neutrino fluxes (either diffuse or from point sources) for neutrino energies above
50 TeV with high efficiency and precision.
In IceCube-Gen2, the very long absorption lengths of the glacial ice at the South Pole will be exploited in order to implement
additional instrumentation with significantly larger string separation distance.

This optimization of the final detector configuration is actively on-going and 
is being modeled by simulating different detector configurations and adjusting the string spacing.
A sample configuration is shown in the upper right panel of Fig.~\ref{fig:1146}.
Note that larger string spacing corresponds to increased
energy threshold and detection efficiency.
The detection of neutrino-induced muon track events scales with the detector instrumentation cross-sectional
area, whereas the neutrino shower events produced by electron and tau flavors and neutral current
interactions scale with the instrumented volume.
For point-source searches, which rely on muon tracks produced by CC interactions
of muon neutrinos in or near the instrumented volume, the sensitivity increases with the projected
cross-sectional area relative to the source direction.
The point-source sensitivities scale approximately with the square-root of the 
increase in the cross-sectional area and linearly with the improvement in angular resolution.
The Gen2 optimized event selections, reconstruction, and analysis methods 
also enhance the sensitivity.
Studies on the angular reconstruction performance, energy
resolution, and veto efficiency for astrophysical events and their refinements 
in the context of point-source and diffuse sensitivity evaluations
are actively ongoing. 
A simulated expected neutrino signal event is shown in the lower plot of Fig.~\ref{fig:1146}. 
For estimated baseline Gen2 performance, which is calculated based on the IceCube performance scaled according to the instrumented volume or cross-sections,
10 years of observation with Gen2 is estimated to be equivalent to more than 200 years of full IceCube observation.

\subsection{KM3NET-ARCA}
\begin{figure*}[!h]
\centering
  \includegraphics[height=2.2in]{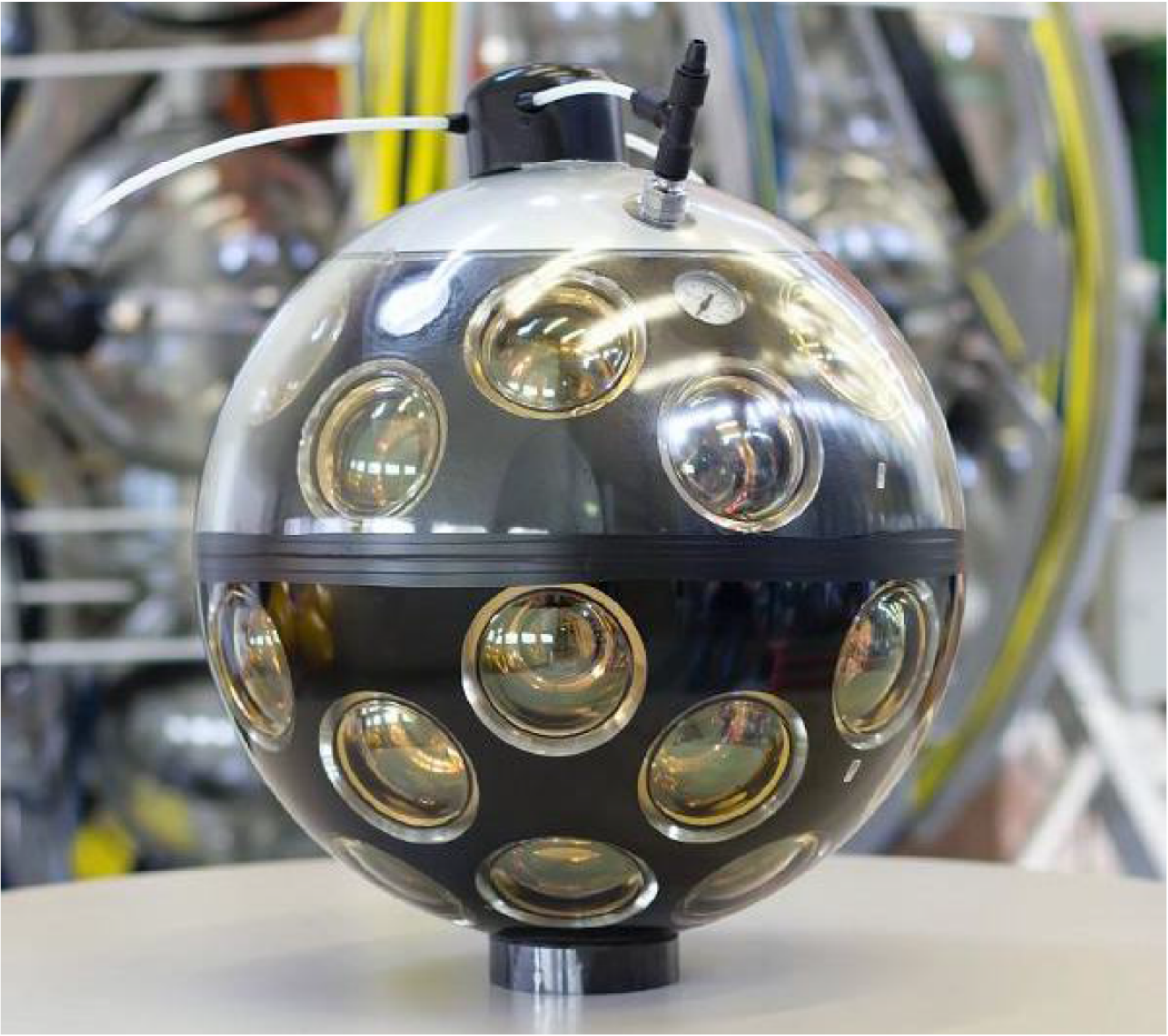} 
  \includegraphics[height=2.in]{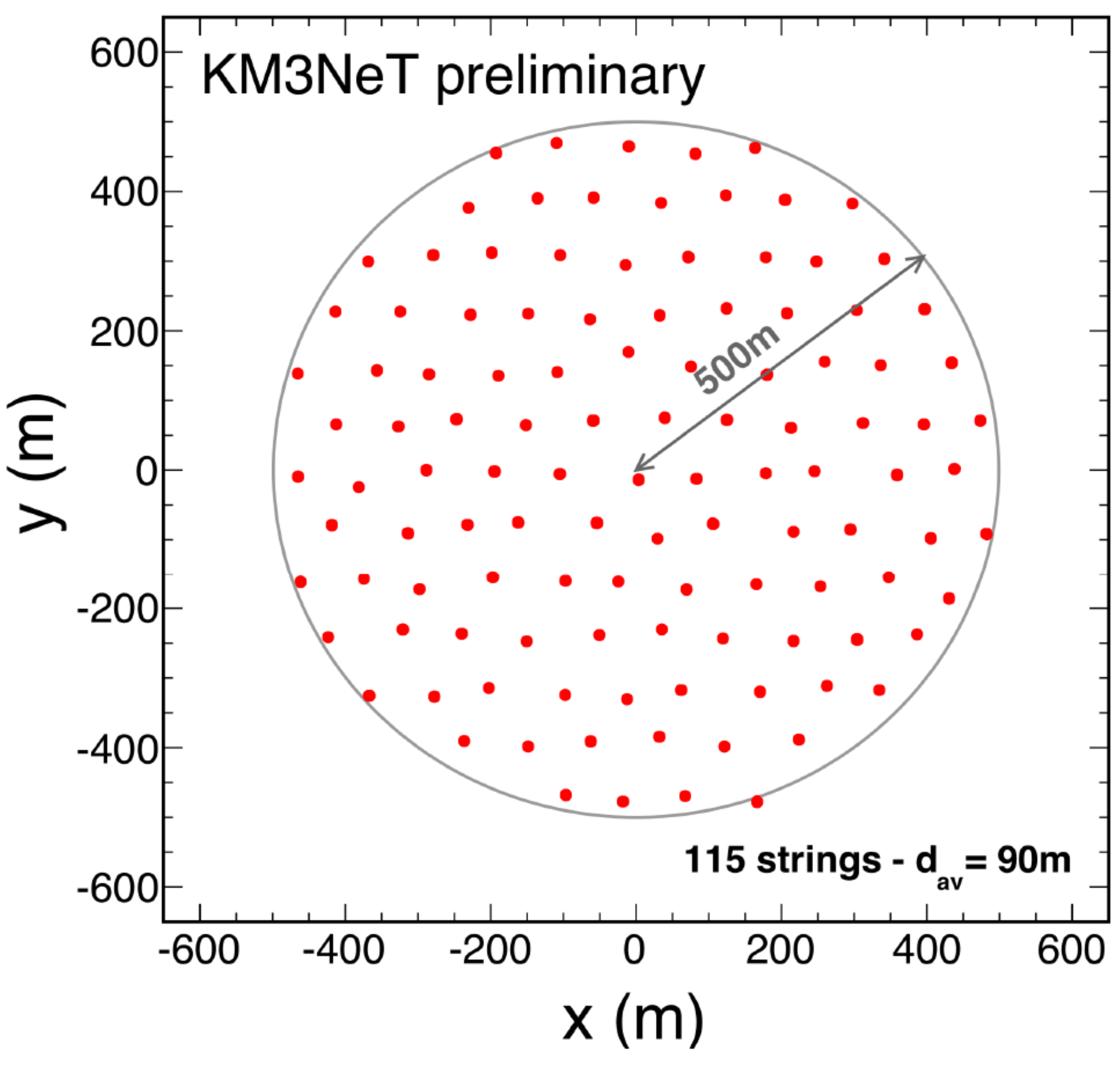} 
  \includegraphics[height=2.in]{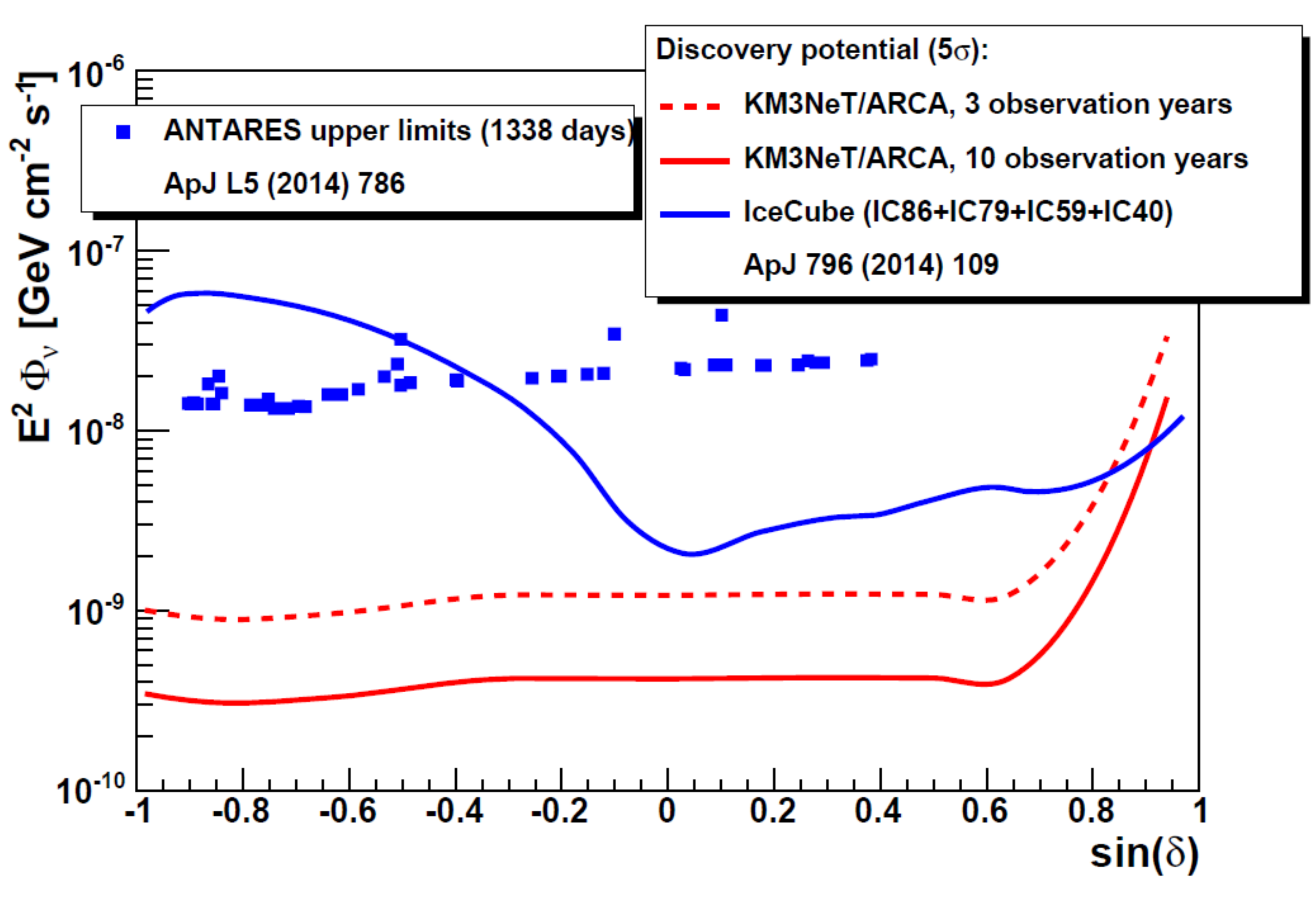} 
\caption{(Upper left) KM3NeT digital optical module.
(Upper right) Footprint of one block of the
KM3NeT/ARCA telescope. The KM3NET2.0-ARCA detector is
composed of two of such blocks.
(Lower) KM3NeT/ARCA 5-sigma
discovery potential per flavor for point sources emitting a
neutrino flux with a E$^{-2}$ spectrum, for three and
10 years of data acquisition. 
}
\label{fig:1158}
\end{figure*}
The KM3NET-ARCA detector that will be constructed at the KM3NeT-It site offshore from Capo Passero (Italy) will be an under-sea Cherenkov neutrino detector located in the northern hemisphere \cite{ICRC2015_1158}.
Because of the advantages offered by its positioning in sea water and in the northern hemisphere,
the KM3NeT/ARCA detector will be a powerful tool for the
exploration and identification of galactic neutrino sources. 
From prior experience with ANTARES, which is also located in sea water, 
the neutrino directions and cascade events will be measured with reasonable angular resolution 
for the performance of point-source investigations. The phase-1 construction, in which a 0.1~km$^3$-scale detector will be constructed, is projected to be completed by 2016. This will be followed by
the phase-2 construction. which will yield a 1~km$^3$-scale detector.
The final phase involves construction of the full KM3NET-ARCA detector, with the aim of implementing two more KM3NET2.0-ARCA-sized detector components.

The basic unit of the KM3NeT detector is the digital optical module (DOM) (Fig.~\ref{fig:1158}), which 
is a pressure-resistant glass sphere of 17 inches in diameter containing 31 3-inch photo-multiplier tubes
(PMTs) with their associated electronics, as well as calibration instrumentation. Nineteen of these
PMTs are located in the lower hemisphere, and are thus oriented downwards, while 12 are positioned in the upper
hemisphere and are, therefore, directed upwards.
These DOMs are installed with a horizontal distance of 90~m and fill a roughly circular
area with a radius of 500~m, as shown in the upper right panel of Fig.~\ref{fig:1158}.
The detector sensitivity for a generic point source with an
E$^{-2}$ spectrum has been calculated as a function of the declination for three years' worth of observation time (Fig.~\ref{fig:1158}). 
The difference in the declination dependence of the sensitivity of the KM3NeT and IceCube detectors is primarily due to the
geographical locations of these detectors.

\subsection{Radio detectors}
Cosmogenic neutrinos, which are induced by interactions between the highest-energy cosmic rays and background
photons in the universe, are expected in the energy region above $\sim$10 PeV.
For proton-dominated UHECR models, the expected flux peaks at EeV energies.
This flux is considered to make a {\it guaranteed} contribution to high-energy neutrino fluxes, as it does not rely on specific neutrino production mechanisms in CR sources. However, even in the simplest case of proton-dominated UHECR models, the neutrino flux depends on the unknown UHECR source redshift distribution function and the maximal energy cut-off of the proton spectra. 
Proton-dominated UHECR models generally provide the most optimistic predictions for cosmogenic neutrino fluxes and are in reach of present neutrino observatories. However, the large experimental uncertainties of the relative contributions of the various elements including UHECR compositions translates into large uncertainties in the GZK neutrino predictions \cite{ahlers2012}. 

\begin{figure*}[!h]
\centering
  \includegraphics[height=1.5in]{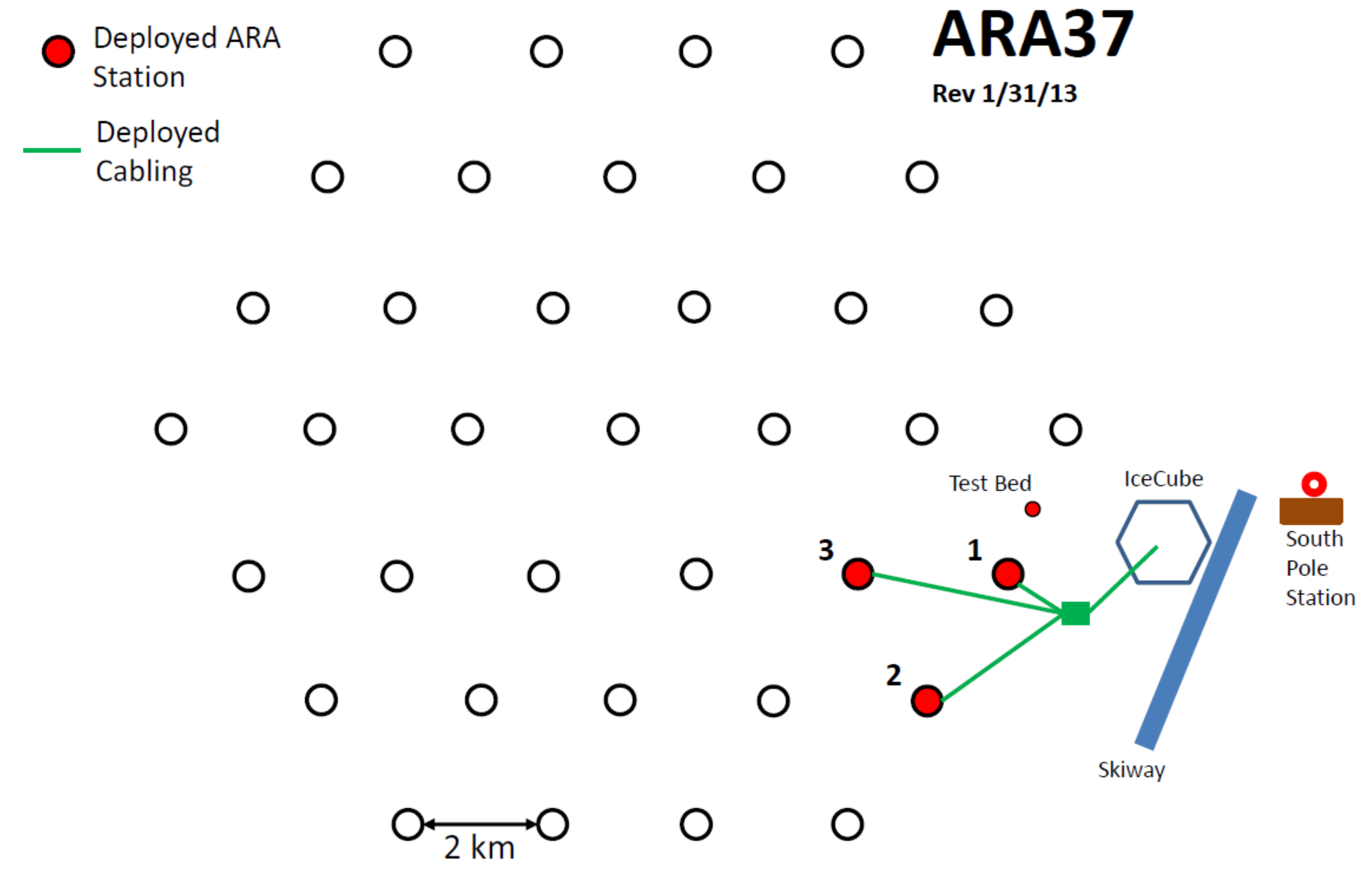} 
  \includegraphics[height=1.8in]{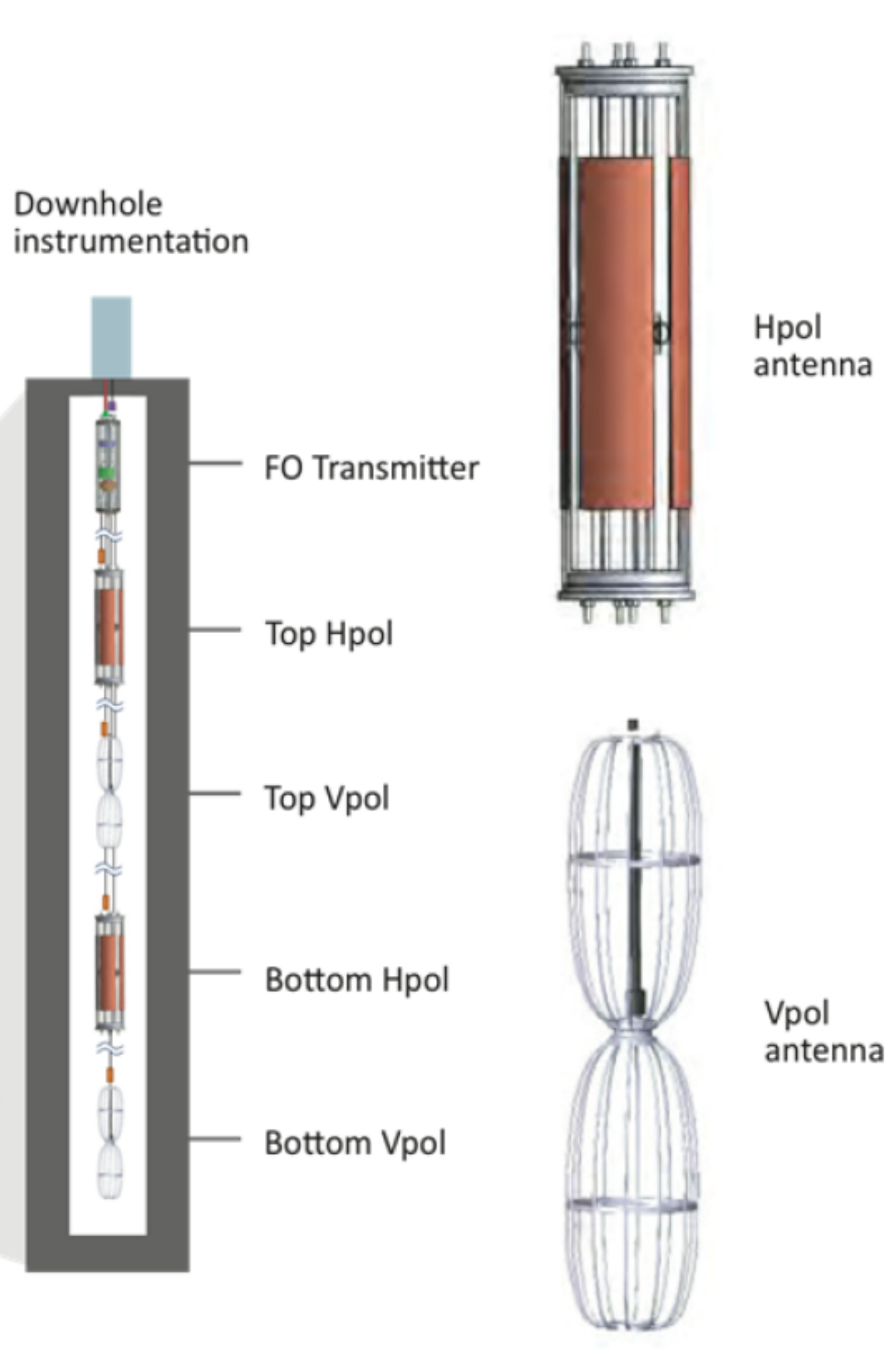} 
  \includegraphics[height=2.5in]{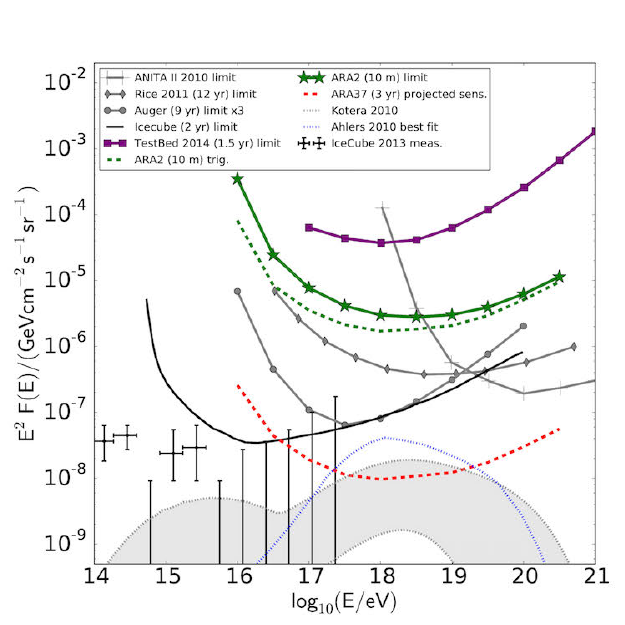} 
\caption{(Upper left) Planned ARA detector configuration at South Pole. The stations are indicated by the black
circles. The red filled circles denote the currently deployed stations.
(Upper right) Design of each hole with a view of the deployed antennas with horizontal and vertical polarizations.
(Lower) Neutrino limits and sensitivities from various detectors including 10-month data analysis of
the two ARA stations.
}
\label{fig:1115}
\end{figure*}
The Askaryan Radio Array (ARA) collaboration has reported the first upper limit on cosmogenic neutrinos \cite{ICRC2015_1115}.
The ARA experiment is a large-scale radio neutrino detector that detects neutrino interactions 
in ice through detection of Cherenkov light in the 200--800-MHz range, with the signal being enhanced 
as a result of the Askaryan effect \cite{ara}. ARA is expected to detect neutrinos above $\sim$10~PeV. Because of the kilometer-scale 
attenuation of radio wavelengths in ice, which is more than an order of magnitude longer than that of optical light, 
only very sparse ($\geq$1~km) detector-station spacing in shallow ice ($\sim$200~m depth) is necessary for ARA.
The ARA detector baseline consists of 37 antenna stations with 2-km spacing in a
hexagonal grid, as shown in Fig.~\ref{fig:1115}. Each station is designed to operate as an autonomous neutrino detector.
Each station comprises 16 measurement antennas, which are deployed in groups of four on strings at the
bottom of 200-m-deep holes. A schematic view of each hole is shown in the upper right panel of Fig.~\ref{fig:1115}.
From analysis using data from two ARA radio antenna stations with an exposure time
of 10~months (Fig.~\ref{fig:1115}) , the upper limits are reported to be 
$3\times10^{-6}$ GeV cm$^{-2}$ s$^{-1}$ sr$^{-1}$
at $10^{18}$~eV.

The Antarctic Ross Ice-Shelf Antenna Neutrino Array (ARIANNA) collaboration plans to construct a neutrino telescope capable of measuring the
diffuse flux of high-energy neutrinos in the $10^8$--$10^{10}$-GeV range \cite{arianna}. 
The ARIANNA Hexagonal
Radio Array (HRA) is a prototype detector intended to test the design performance
of the full ARIANNA telescope. During the 2014--2015 austral summer, the HRA installation was completed at the ARIANNA detector site on the
Ross Ice Shelf of Antarctica \cite{ICRC2015_1087}. This facility consists of eight independent detector stations, each of which
has four log-periodic dipole antennas and an autonomous data acquisition (DAQ) system. Local solar power is used and 
the stations measure radio pulses in the 50-MHz--1-GHz frequency range.
All eight stations were installed by early December 2014 and were in operation until the sun began to set
in early April 2015.
Special calibration data was collected and the radio pulse arrival direction
resolution  was estimated. This was confirmed using the calibration data.
\begin{figure*}[!h]
\centering
  \includegraphics[height=2.in]{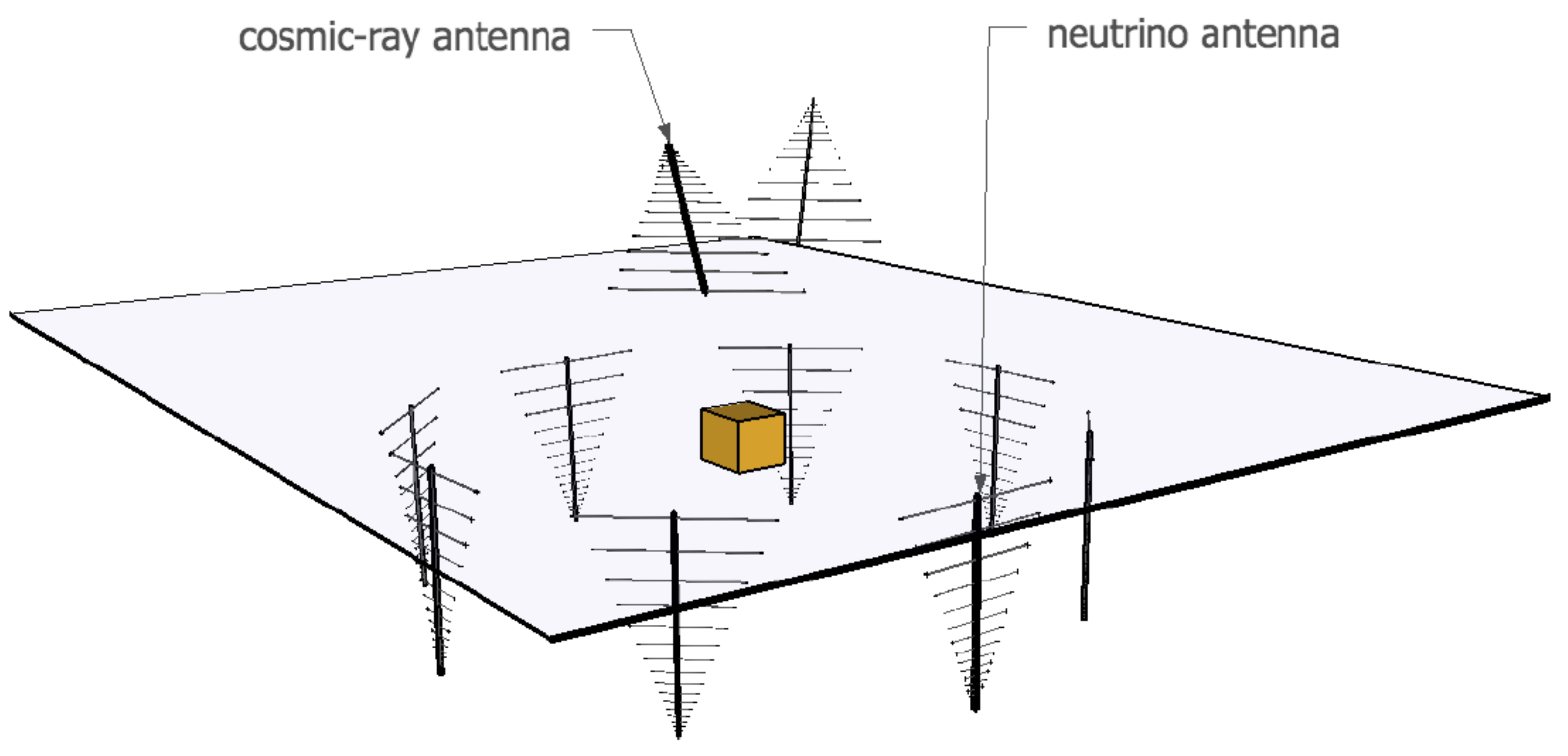} 
\caption{Schematic view of ARIANNA-HEX configuration.}
\label{fig:}
\end{figure*}

Furthermore, a report on ongoing work regarding a site survey in Greenland was presented at ICRC 2015 \cite{ICRC2015_1150}.
This survey is being performed to explore the radio-frequency environment of the Summit Station 
in Greenland and to test several potential configurations for a phased dipole array.
The Greenland Neutrino Observatory (GNO) collaboration calculated an attenuation length depth profile from 
the depth-averaged attenuation length using measured temperature profiles of the Summit Station
and the measured temperature dependence of the attenuation length. The results indicate that
the  Summit Station sites are also suitable locations for the construction of radio neutrino detectors.

\begin{figure*}[!h]
\centering
  \includegraphics[height=2.in]{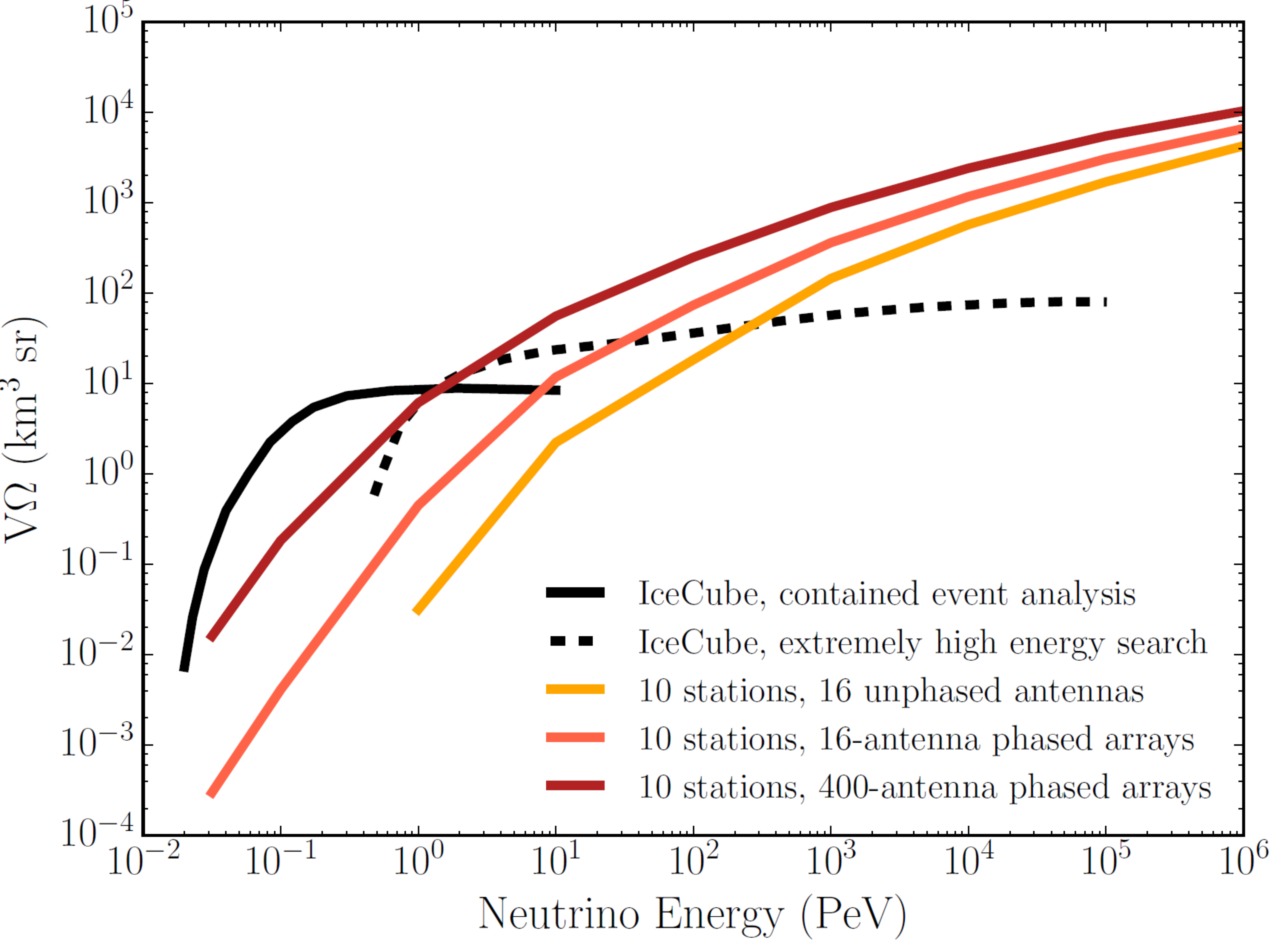} 
\caption{Comparison of neutrino effective area with and without implementation of phased array method.
With the phased array, the energy threshold can be reduced significantly.}
\label{fig:1171}
\end{figure*}

The GNO site survey included a test of a concept to reduce the neutrino energy threshold 
using radio antenna arrays \cite{ICRC2015_1171}.
Through coherent summation of the waveforms from each antenna, the signal amplitude increases linearly
and the noise grows in quadrature, resulting in an increase in the signal-to-noise-ratio.
Implementation of this phased array method in either Greenland or Antarctica has the potential to both improve 
sensitivity to GZK neutrinos and potentially reduce the energy
threshold of the radio arrays, possibly down to the PeV scale. This will facilitate measurement of the
PeV-energy neutrinos being observed with the radio Cherenkov technique, as shown in Fig.~\ref{fig:1171}.

\begin{figure*}[!h]
\centering
  \includegraphics[height=2.in]{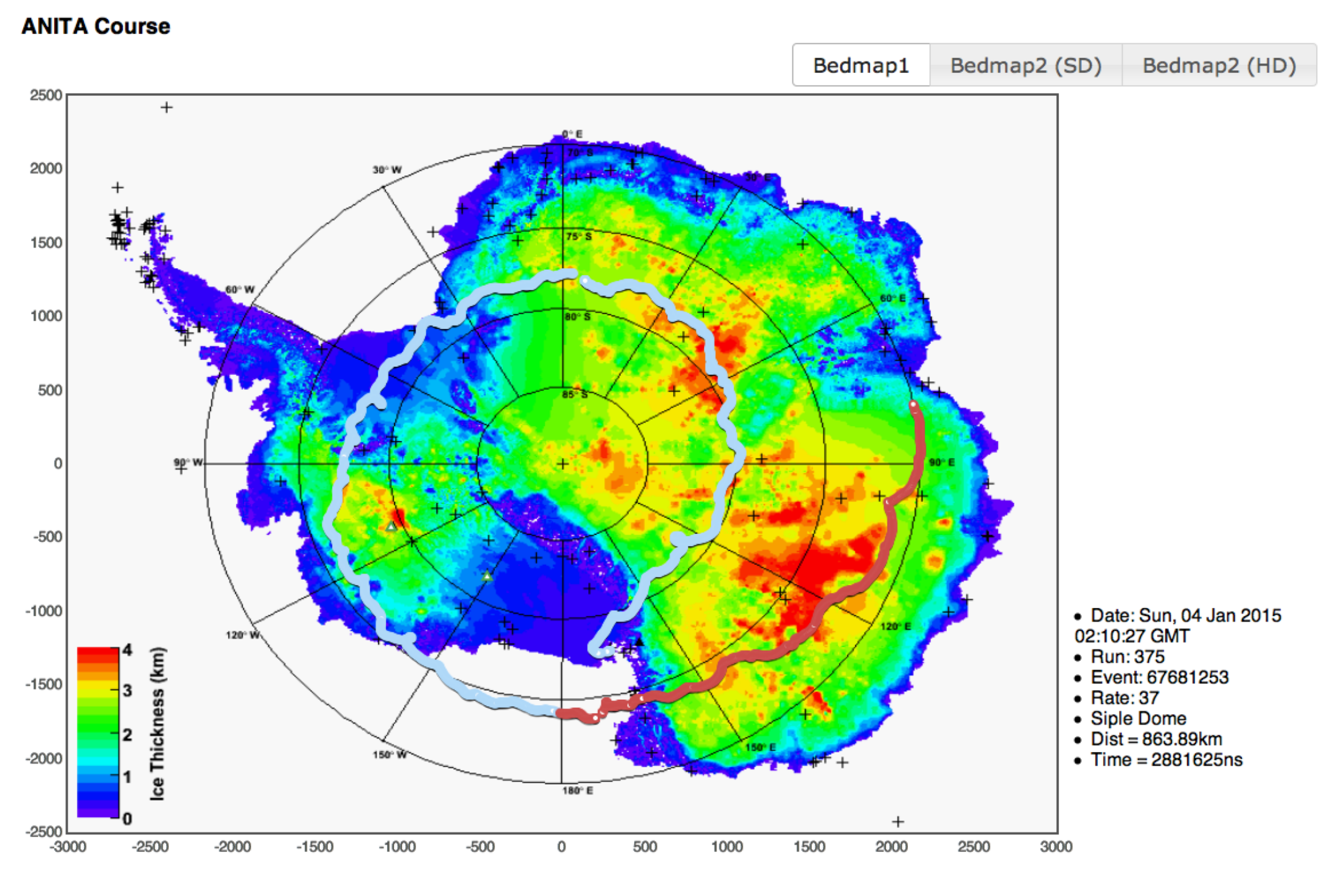} 
\caption{ANITA flight path on ice thickness map.}
\label{fig:1111}
\end{figure*}

The Antarctic Impulsive Transient Antenna (ANITA) is a long-duration balloon 
mission with the goal of measuring the energy spectra of cosmic UHE neutrinos.
The extra-galactic neutrino flux that ANITA seeks to measure peaks in the EeV energy range. The ANITA-III flight report was presented at the ICRC 2015 conference \cite{ICRC2015_1111}.
ANITA-III was the third flight in the series of ANITA missions. The payload was launched on 2014-12-17 at 21:24 UT 
and the flight was terminated on 2015-01-08 at approximately 23:50 UT after 22 days afloat. 
This is a relatively short flight compared to the other two
ANITA flights conducted to date. Although the payload was performing well and was approved for additional
float time, the stratospheric trajectory was spiraling off the Antarctic continent. Thus, it was 
decided to terminate the flight while the balloon was near a reasonable recovery site, in order to minimize the possibility of payload loss. The final landing place was approximately 100 nm from Davis Station (AU), on the ice sheet and at an elevation
of approximately 2500 m. Figure~\ref{fig:1111} shows a map of the trajectory.

\begin{figure*}[!h]
\centering
  \includegraphics[height=2.in]{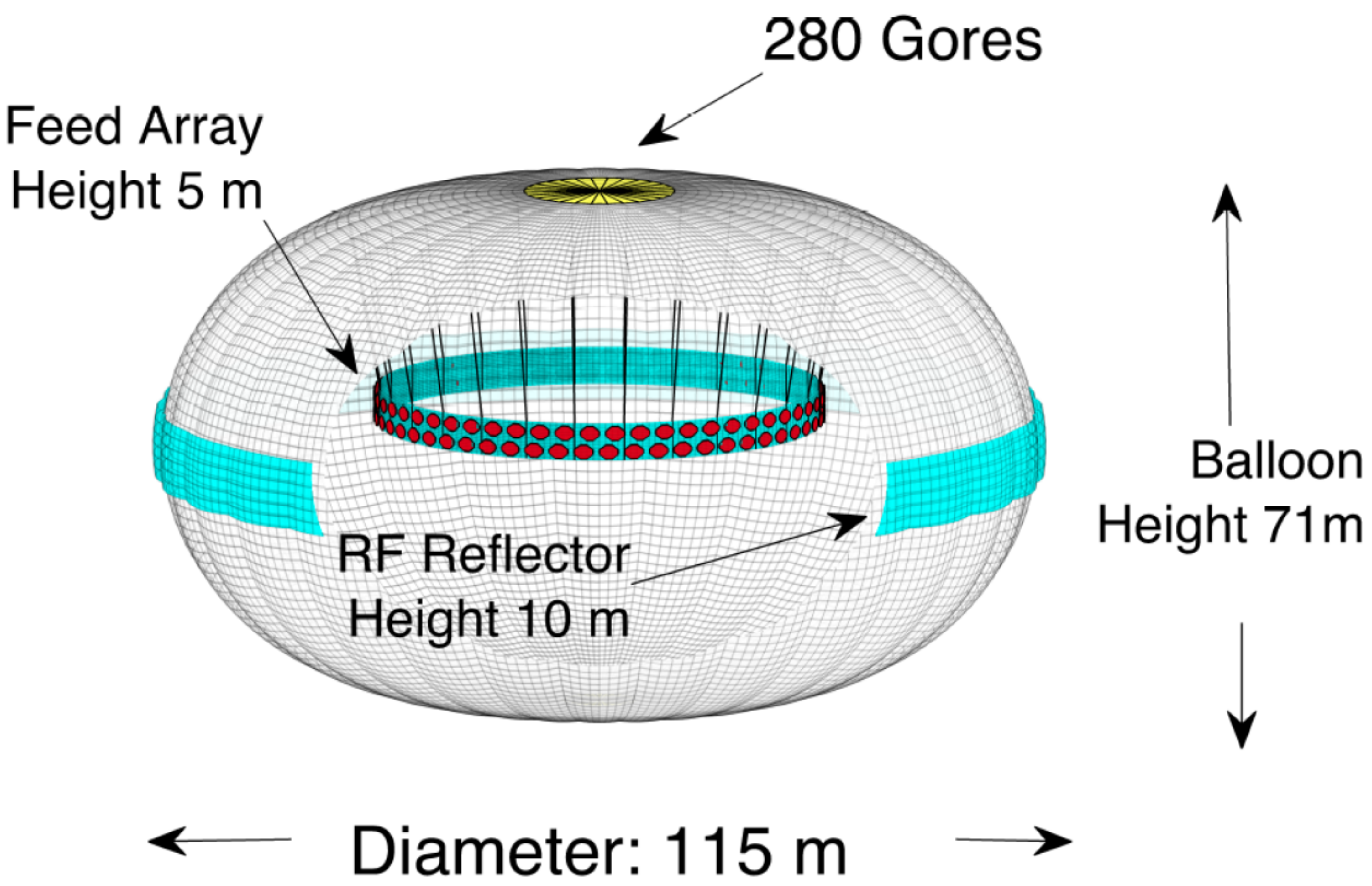} 
  \includegraphics[height=2.in]{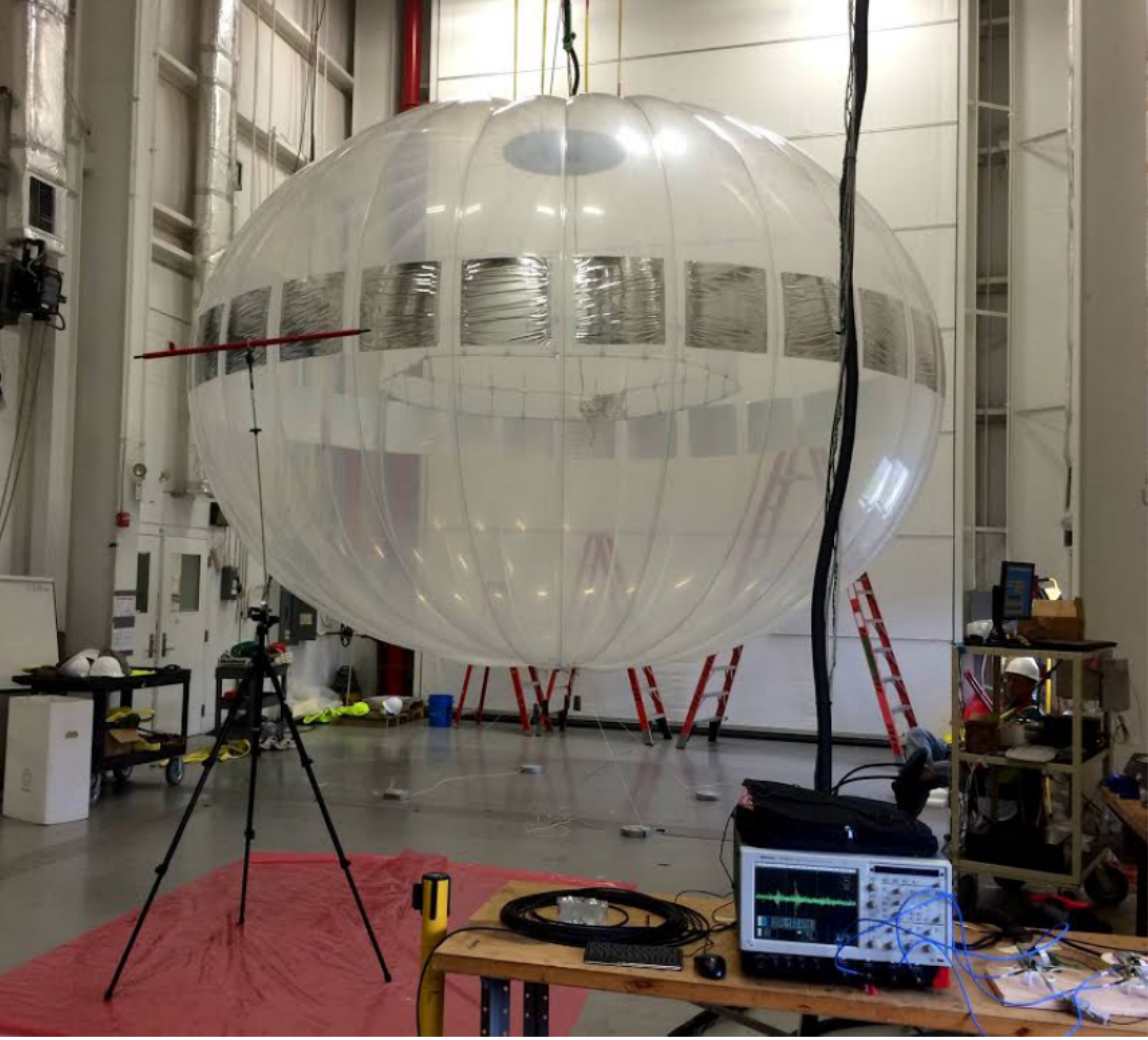} 
\caption{Schematic model of EVA full-scale 18.7-Mcft super-pressure balloon. 
The reflector panels (light blue) are cut-away in the image front center, so as to reveal the internal
feed array.
(Right) Fully deployed 20-ft-diameter balloon, with DAQ system in foreground. 
}
\label{fig:1151}
\end{figure*}

A new balloon antenna design for the Exavolt Antenna (EVA) project is in the prototype development phase \cite{ICRC2015_1151}.
EVA aims to implement a 30-dBi-gain antenna, exploiting the surface of a super-pressure balloon (SPB) to 
improve sensitivity to neutrino-induced radio-impulsive transients by a factor of 100 over that of ANITA.
The EVA design is shown in the left panel of Fig.~\ref{fig:1151}.
In September 2014, the EVA collaboration performed a hangar test using a smaller-scale 
prototype, as shown in the right panel of Fig.~\ref{fig:1151}.
This balloon included 50-cm-high aluminized-mylar reflective panels
and, prior to inflation of the balloon, a scale-model of the antenna feed array membrane
was inserted through the top end-plate of the balloon.
The feed array membrane was instrumented with
dual-polarization sinuous patch antennas over a portion of its circumference, and four of these
antennas were also instrumented with microwave receivers coupled to radio-frequency (RF)-over-fiber transceivers,
which supplied the optical fiber with RF signals.

%
\section{Summary}
Following the first observation of astrophysical neutrino flux by IceCube, 
the field of neutrino astronomy has become increasingly active.
The IceCube detector has been in stable operation and has now reached the critical phase of increasing statistics
to further measure the properties of cosmic neutrinos.
The ANTARES detectors are also in stable operation. While their size remains limited,
these detectors have exhibited good sensitivity to galactic neutrino point sources.
Motivated by the success of the 1-km$^3$-scale Cherenkov neutrino detector and the importance of the
physical topics that may be addressed, new 10-km$^3$ detector designs are being actively studied for implementation in 
Antarctica and the Mediterranean Sea. It is important to improve detector design so that cosmic-ray origins can be identified through
observation of neutrino point sources.
It is also important to measure the neutrino flux above 10~PeV. The current most realistic strategy is observation of cosmogenic neutrinos through the 
detection of Cherenkov radio emission via the Askaryan effect, so as to determine the origin of the highest-energy cosmic rays.
\FloatBarrier
\bibliographystyle{JHEP}
\bibliography{Ishihara_Rapporter}
\end{document}